\documentclass[referee]{raa}

\usepackage{graphicx,times}
\usepackage{natbib}
\usepackage[scheme=plain]{ctex}
\usepackage{amssymb,amsmath}
\usepackage{color}
\usepackage{booktabs}
\usepackage{multirow}
\bibpunct{(}{)}{;}{a}{}{,}

\usepackage{xcolor} 
\usepackage[pagebackref=true]{hyperref}

\begin{document}

\title{Particle Swarm Optimization Strategies for Lensed Gravitational Waves: Parameter Estimation, Source Classification, and Accelerated Sampling}
%\title{A Particle Swarm Optimization Framework for Lensed Gravitational Wave Data Analysis}
\volnopage{ {\bf 20XX} Vol.\ {\bf X} No. {\bf XX}, 000--000}
\setcounter{page}{1}

\author{Wei-Qing Yu  %% Put your Chinese name in "( )" if you like.
	\inst{1}, Meng Zhang \inst{1,*}, Xiaobo Zou\inst{2}, Xiao Guo\inst{3,4,*}\footnotetext{$*$Corresponding Authors.}, 
}
%% Here is an example of three authors come from different institutes.
%% For single author or all the authors from an institute, use "\inst{}" only

\institute{School of Cyber Science and Engineering, Qufu Normal University, Qufu 273165, China; {\it zhangmeng@qfnu.edu.cn}\\
%% Please give the E-mail address of the author, to whom future correspondence and
%% offprint requests will be sent.
\and
School of Fundamental Physics and Mathematical Sciences, Hangzhou Institute for Advanced Study, University of Chinese Academy of Sciences, No.1 Xiangshan Branch, Hangzhou 310024, China
\and
School of Physics and Technology, Wuhan University, Wuhan 430072, China ; {\it guoxiao17@mails.ucas.ac.cn}
\and
Institute for Gravitational Wave Astronomy, Henan Academy of Sciences, Zhengzhou 450046, Henan, China
\\
\vs \no
{\small Received 20XX Month Day; accepted 20XX Month Day}
}

\abstract{The detection and parameter estimation of lensed gravitational wave (GW) signals are crucial for astrophysics and cosmology. Strong lensing introduces extra parameters, rendering the search high-dimensional, multi-modal, and strongly nonlinear. We apply Particle Swarm Optimization (PSO) to the search, parameter estimation, and classification of lensed compact binary coalescence GWs within a four-detector coherent network. Using the restricted 2-PN waveform as a template and a model-independent double-image lens parameterization, we assess PSO on both model-matched 2-PN signals and template-mismatched IMRPhenomD signals. Results show effective recovery of chirp mass, mass ratio, sky location, and lens parameters in the 2-PN case, with mass biases appearing under waveform mismatch. A three-class classification pipeline combining 12 PSO runs, majority voting, and a flux ratio threshold achieves 86.19\% accuracy in distinguishing lensed GWs, unlensed GWs, and noise. We further propose PSO\_GAUSS\_MIX method, which constructs a Gaussian mixture prior from PSO estimates for nested sampling, yielding speedups of $\sim1.38\times$ in the restricted 2-PN cases while preserving posterior accuracy. These results establish PSO as an efficient optimizer for lensed GW searches and an effective prior constructor for Bayesian parameter estimation.
%The detection and parameter estimation of lensed gravitational wave (GW) signals are important for astrophysics and cosmology. Lensing introduces additional parameters, making the search high-dimensional, multi-modal, and strongly nonlinear. We apply Particle Swarm Optimization (PSO) to search, parameter estimation, and signal classification of lensed compact binary coalescence GWs under a four-detector coherent network. Using the restricted 2-PN waveform as template and a model-independent double-image lens parameterization, we evaluate PSO on both model-matched 2-PN and template-mismatched IMRPhenomD signals. Results show effective recovery of chirp mass, mass ratio, sky location, and lens parameters in the 2-PN case, with mass biases under mismatch. A three-class classification pipeline using 12 PSO runs, majority voting, and a flux ratio threshold achieves $86.19\%$ accuracy in distinguishing lensed GWs, unlensed GWs, and noise. We further propose PSO\_GAUSS\_MIX, which constructs a Gaussian mixture prior from PSO estimates for nested sampling , achieving $\sim 2.84\times$ and $2.07\times$ speedups in the two cases while preserving posterior accuracy. These results establish PSO as an efficient optimizer for lensed GW searches and a viable prior constructor for Bayesian parameter estimation.
}

\authorrunning{Yu et al. }            %author_head in even pages
\titlerunning{PSO for lensed GW }  % title_head in odd pages
\maketitle

%_____________________________ sections below_________________________%
\section{Introduction}           
\label{sect:intro}
Since the Laser Interferometer Gravitational Wave Observatory (LIGO) made the first direct detection of GW150914, a gravitational wave (GW) signal from a binary black hole merger, in 2015 (\citealt{Abbott+etal+2016}), GW astronomy has advanced rapidly. GWs are a fundamental prediction of Einstein’s general theory of relativity, and the detection of GW150914 both directly confirmed this prediction and opened a new observational window for astronomy and cosmology \citep{Diego+etal+2021,Abbott+2017}. Before the release of GWTC-5.0, the LIGO-Virgo-KAGRA (LVK) Collaboration had cataloged 218 GW events \citep{Abac+etal+2025a}. The first three observing runs yielded 90 detections (\citealt{Abbott+etal+2023}), and O4a contributed 128 additional events, more than doubling the catalog size; additional candidates remained under analysis (\citealt{Abac+etal+2025b}). On 26 May 2026, the LVK Collaboration released GWTC-5.0, which added 161 events from O4b and raised the total number of confirmed GW detections to 390; together, O4a and O4b account for approximately 75\% of all cataloged GW events to date (\citealt{LVK+2026}). As detector sensitivity continues to improve and next-generation facilities such as the Einstein Telescope and Cosmic Explorer come online, the GW detection rate is expected to increase substantially (\citealt{Punturo+etal+2010,Maggiore+etal+2020}).

Gravitational lensing is also a fundamental prediction of general relativity (\citealt{Bartelmann+2010}). When a GW passes near a massive foreground lens, the curvature of spacetime can deflect its trajectory and produce multiple images that reach the detector at different times (\citealt{Abbott+etal+2023,Sereno+etal+2010}). Lensed GWs could therefore serve as powerful probes in GW astronomy and cosmology (\citealt{Grespan+Biesiada+2023,Abbott+etal+2024}). Some astrophysical objects and structures, including intermediate-mass black holes (\citealt{Meena+2024,Gais+etal+2022,Lai+etal+2018}) and dark matter halos (\citealt{Guo+Lu+2022,Choi+etal+2021}), remain challenging to detect through conventional electromagnetic observations. A massive foreground lens can magnify a GW signal, introduce relative time delays, generate multiple images, and imprint distortions on the observed waveform (\citealt{Dai+Venumadhav+2017,Oguri+2018}). Detecting and characterizing such lensed GWs may therefore offer direct insight into compact lenses and the nature of dark matter (\citealt{Urrutia+Vaskonen+2022}).

In recent years, extensive searches for GW lensing signatures have been carried out using both template-based methods, including matched filtering (\citealt{Jaranowski+Krolak+2012,Janquart+etal+2023,Barsode+etal+2025}), and deep-learning approaches (\citealt{Wang+etal+2020,Schafer+etal+2023,Goyal+etal+2021}). The LVK Collaboration has also performed systematic analyses of third-observing-run data to search for lensed GWs (\citealt{Abbott+etal+2021,Abbott+etal+2024}). Although no lensed GW has yet been conclusively confirmed, current-generation detectors may still yield the first such detection in upcoming higher-sensitivity runs (\citealt{Abbott+etal+2024,Barsode+etal+2025}).

However, detecting and characterizing lensed GWs is challenging. First, a typical compact-binary coalescence is described by 15 or more parameters (\citealt{RomeroShaw+etal+2020}), and lensing adds at least two more, further enlarging the search space. This increase in dimensionality makes grid-based matched filtering susceptible to the ``curse of dimensionality'', and the computational cost can increase exponentially with the number of parameters (\citealt{Tiglio+Villanueva+2022}). Second, the likelihood surface for GW signals is often highly multimodal and may contain numerous local optima, which can hinder optimization algorithms from identifying the global solution \citet{Ashton+etal+2021}. In addition, the underlying lens model is often unknown. Different models, such as point-mass and singular isothermal sphere lenses, can produce distinct waveform morphologies, further complicating matched-filter searches (\citealt{Janquart+etal+2021}). Efficient and robust parameter estimation in such a high-dimensional, multimodal setting therefore remains a central challenge in research on lensed GWs.

Particle swarm optimization (PSO) is a stochastic global optimization algorithm (\citealt{Kennedy+Eberhart+1995}) that has been widely used in GW signal searches and data analysis. For example, Wang and Mohanty first applied PSO to compact-binary coalescence (CBC) searches using data from a single detector (\citealt{Wang+Mohanty+2010}). Normandin subsequently extended this framework to fully coherent network parameter estimation using locally optimal PSO (\citealt{Weerathunga+Mohanty+2017,Normandin+etal+2018}), This approach later achieved near-real-time performance (\citealt{Normandin+Mohanty+2020}). More recently, PSO has been applied to searches for multi-target Galactic binary GWs (\citealt{Zhang+etal+2021,Gao+etal+2024}) and extreme-mass-ratio inspiral signals (\citealt{Zou+etal+2024,Zou+Mohanty+2024,Zou+etal+2025}). Despite these advances, the application of PSO to lensed GW searches and parameter estimation remains underexplored.

We investigate the effectiveness of PSO for fully coherent all-sky CBC searches. Following \citet{Weerathunga+Mohanty+2017}, we construct a four-detector GW
network and use the aLIGO design-sensitivity power spectral density (PSD) to simulate detector noise (\citealt{LALSuite+2020}). For signal modeling, we use the 2-PN binary-inspiral waveform as the GW template and incorporate a model-independent double-image lensing model. We then use PSO to search the extended parameter space and evaluate its performance in signal detection and parameter recovery using the generalized likelihood ratio test (GLRT) and maximum likelihood estimation (MLE)(\citealt{Kay+1993}).

The remainder of this paper is organized as follows. Section~\hyperref[sect:Obs]{2} describes the data and signal models, and Section~\hyperref[sect:PSO]{3}
introduces the PSO algorithm and the selection of its parameters. Section~\hyperref[sect:Parameter recovery]{4} reports the PSO-based point-estimation results,
Section~\hyperref[sect:classification]{5} presents the results of a voting-based classification method, and Section~\hyperref[sect:psoGaussMix]{6} describes the
method used to reconstruct posterior distributions and presents the corresponding results. Finally, Section~\hyperref[sect:conclusion]{7} summarizes the main
findings and discusses future prospects.
% 本文其余部分安排如下。第~\hyperref[sect:Obs]{2}节描述数据和信号模型，第~\hyperref[sect:PSO]{3}节介绍PSO算法及其参数选择。第~\hyperref[sect:Parameter recovery]{4}节报告基于PSO的点估计结果，第~\hyperref[sect:classification]{5}节展示基于投票机制的分类方法结果，第~\hyperref[sect:psoGaussMix]{6}节描述用于重建后验分布的方法并给出相应结果。最后，第~\hyperref[sect:conclusion]{7}节总结主要发现并讨论未来前景。

\section{DATA and Signal Models}
\label{sect:Obs}
%%==========================2.1探测器系数===============================
\subsection{Detector network and response}
We adopt a detection network consisting of four GW detectors: LIGO Hanford (H1), LIGO Livingston (L1), Virgo (V1), and KAGRA (K1) (\citealt{Srivastava+etal+2018}). The position vector $\boldsymbol{r}^J$ and arm-orientation unit vectors $(\boldsymbol{\hat{u}}^J,\boldsymbol{\hat{v}}^J)$ of each detector in the Earth-centered Earth-fixed (ECEF) frame are taken from \citet{Weerathunga+Mohanty+2017}; their components are listed in Table~\ref{ECEF}. The unit vector $\boldsymbol{\hat{n}}$ pointing from the ECEF origin to the GW source is defined as
% 我们采用由四个引力波探测器组成的探测网络：LIGO Hanford（H1）、LIGOLivingston（L1）、Virgo（V1）和KAGRA（K1）（\citealt{Srivastava+etal+2018}）。每个探测器在地心地固坐标系（ECEF）中的位置矢量$\boldsymbol{r}^I$和臂取向单位矢量$(\boldsymbol{\hat{u}}^I,\boldsymbol{\hat{v}}^I)$取自\citealt{Weerathunga+Mohanty+2017}；其各分量列于表~\ref{ECEF}。从ECEF原点指向引力波源的单位矢量$\boldsymbol{\hat{n}}$定义为

\begin{equation}
    \boldsymbol{\hat{n}} = (\cos\alpha\cos\delta,\sin{\alpha}\cos\delta,\sin\delta)
\end{equation}
where $\alpha$ and $\delta$ represent the right ascension and declination, respectively, and $\psi$ is the polarization angle. The time delay of the signal arriving at the $J$th detector relative to the ECEF origin is
% 其中，$\alpha$和$\delta$分别代表赤经和赤纬，$\psi$为极化角。信号到达第$I$个探测器相对于ECEF原点的时间延迟为
\begin{equation}
  \Delta^J =\frac{\boldsymbol{\vec{r}^J}\cdot \boldsymbol{\hat{n}}}{c}  
\end{equation}
The antenna response of a detector is determined by the double contraction between the detector tensor and the GW polarization-basis tensor. The detector tensor of the $J$-th detector is defined as
% 探测器的天线响应由探测器张量与引力波极化基张量的双重收缩确定，第$I$个探测器的探测器张量定义为
\begin{equation}
    % \overleftrightarrow{d^I} = \frac{1}{2}(\hat{u}^I\otimes \hat{u}^I - \hat{v}^I \otimes \hat{v}^I) \\
    d^J_{ab} = \frac{1}{2}(\hat{u}^J_a \otimes \hat{u}^J_b - \hat{v}^J_a \otimes\hat{v}^J_b)
\end{equation}
where $a,b=1,2,3$.
The two tensor polarization bases of a GW can be written as (~\citealt{Whelan+2013})
% GW的两种张量极化基可写为(~\citealt{Whelan+2013})
\begin{align}
    % \overleftrightarrow{\epsilon}_+ = \boldsymbol{e^R_x} \otimes \boldsymbol{e^R_x} - \boldsymbol{e^R_y} \otimes \boldsymbol{e^R_y} \\
    % \overleftrightarrow{\epsilon}_\times = \boldsymbol{e^R_x} \otimes \boldsymbol{e^R_y} + \boldsymbol{e^R_y} \otimes \boldsymbol{e^R_x}\\
    \epsilon_+^{ab} &= e_x^a\otimes e_x^b - e_y^a \otimes e_y^b \\
    \epsilon_\times^{ab} &= e_x^a \otimes e_y^b + e_y^a \otimes e_x^b
\end{align}
where $(e^a_x,e^a_y)$ is a pair of orthogonal reference basis vectors in the plane perpendicular to $\boldsymbol{\hat{n}}$, defined as
% 其中，$(\boldsymbol{e^R_x},\boldsymbol{e^R_y})$为垂直于$\hat{n}$的平面内一组正交参考基矢量，其定义为
\begin{align}
    e_x^a &= (\sin\alpha,\,-\cos\alpha,\,0) \\
    e_y^a &= (-\cos\alpha\sin\delta,\,-\sin\alpha\sin\delta,\,\cos\delta)\\
    % \boldsymbol{e^R_x} &= (\sin\alpha,-\cos\alpha,0 )  \\
    % \boldsymbol{e^R_y} &= (-\cos\alpha\sin\delta,-\sin\alpha\sin\delta,\cos\delta)
\end{align}
Thus, the sky-response coefficients $U^J_+$ and $U_\times^J$ are defined as
% 由此，天空响应系数$U^I_+$和$U_\times^I$定义为
\begin{equation}
    % U^I_+ = \overleftrightarrow{\epsilon_+} :\overleftrightarrow{d^I}   \qquad 
    % U_\times^I = \overleftrightarrow{\epsilon_\times} : \overleftrightarrow{d^I} \\
    U^J_+ = \epsilon_+^{ab} d^J_{ab}, \qquad
    U_\times^J = \epsilon_\times^{ab}d^J_{ab}.
\end{equation}
%where the symbol ``:'' denotes the double-dot product operation of tensors.
Because $U_+^J$ and $U_\times^J$ depend only on the sky position $(\alpha,\delta)$, we analytically expand the antenna response patterns as trigonometric functions of these coordinates during preprocessing. The expansion coefficients are precomputed and stored for direct lookup, thereby avoiding redundant detector-tensor contractions and reducing computational overhead. The sky-response functions $F^J_+$ and $F^J_\times$ are then obtained by rotating $U^J_+$ and $U^J_\times$ by the polarization angle $\psi$:
% 由于$U_+^I$和$U_\times^I$仅依赖于天空位置$(\alpha,\delta)$，我们在预处理阶段将天线响应模式解析展开为这些坐标的三角函数。展开系数被预先计算并存储，以便直接查找，从而避免重复的探测器张量收缩并降低计算开销。随后，通过极化角$\psi$对$U^I_+$和$U^I_\times$进行旋转，可得到天空响应函数$F^I_+$和$F^I_\times$：
\begin{equation}
    \begin{pmatrix}
        F_+^J \\ F_\times^J 
    \end{pmatrix} 
    =
    \begin{pmatrix}
        \cos 2\psi & \sin 2\psi \\ - \sin 2\psi & \cos 2\psi
    \end{pmatrix} 
    \begin{pmatrix}
        U_+^J \\ U_\times^J
    \end{pmatrix}
\end{equation}
Finally, the strain response of the $J$th detector is \citep{Schutz+2011}
% 最终，第$I$个探测器的应变响应为(~\citealt{Schutz+2011})
\begin{equation}
    h^J(t) = F^J_+(\alpha,\delta,\psi)h_+(t-\Delta^J) +  F^J_{\times}(\alpha,\delta,\psi)h_\times(t-\Delta^J)
\end{equation}

\begin{table}
	\bc
	\begin{minipage}[]{100mm}
		\caption[]{Components of the detector position vectors and arm direction unit vectors in the Earth-centered Earth-fixed (ECEF) frame \label{ECEF}}
		\end{minipage}
	\setlength{\tabcolsep}{8pt}
	\small
	\begin{tabular}{ccccc}
		\hline\noalign{\smallskip}
		  &  H1               & L1       & V1 & K1    \\
		\hline\noalign{\smallskip}
		$u_x$ & -0.2239 & -0.9546     & -0.7005 &-0.4300  \\
		$u_y$ & 0.7997 & -0.1416     & 0.2085 & -0.8363 \\
		$u_z$ & 0.5569 & -0.2622     & 0.6826 & 0.3400 \\
		$v_x$ & -0.9140 & 0.2977     & -0.0538 & 0.6821  \\
		$v_y$ & 0.0261 & -0.4879     & -0.9691 & -0.0542  \\
		$v_z$ & -0.4049 & -0.8205     & 0.2408 & 0.7292  \\
		$x$ & $-2.1614\times 10^6$ & $-7.4276\times 10^4$     & $4.5464\times 10^6$ & $-3.7769\times 10^6$  \\
	    $y$ & $-3.8347\times 10^6$ & $-5.4963\times 10^6$     & $8.4299\times 10^5$ & $3.4839\times 10^6$ \\
		$z$ & $4.6004\times 10^6$ & $3.2243\times 10^6$     & $4.3786\times 10^6$ & $3.7667\times 10^6$  \\
		\noalign{\smallskip}\hline
	\end{tabular}
	\ec
	%% place \tablecomments and \tablerefs below \end{center| and \end{center}:
	%% you may leave the table-width parameter to editors or set to your actual size
	\tablecomments{0.86\textwidth}{
	In the table, $(u_x,u_y,u_z)$ and $(v_x,v_y,v_z)$ denote the components of the unit direction vectors of the two interferometer arms of the detector in the ECEF frame, and $(x,y,z)$ denote the components of the detector vertex position vector in the ECEF frame.The unit of the position vector components is m, and the arm direction vectors are dimensionless unit vectors.}
\end{table}
%%=================================================2-PN===================================================
\subsection{Restricted 2-PN signal}
In what follows, $a(t)$ denotes a time-domain function, and $\tilde{a}(f)$ denotes its Fourier transform in the frequency domain. We adopt the restricted 2-PN
approximation as the GW template for nonspinning compact-binary systems. In this approximation, the signal phase is expanded and retained up to order $(v/c)^4$,
whereas the amplitude is kept only at leading order (\citealt{Blanchet+etal+1995}).
% 在下文中，$a(t)$表示时域函数，$\tilde{a}(f)$表示其在频域中的傅里叶变换。我们采用受限2-PN近似作为非自旋致密双星系统的引力波模板。在该近似下，信号相位被展开并保留至$(v/c)^4$阶，而振幅仅保留主阶项（\citealt{Blanchet+etal+1995}）。

In the restricted 2-PN framework, the two orthogonal polarization waveforms in the frequency domain, $\tilde{h}_+(f)$ and $\tilde{h}_\times(f)$, are expressed as
(\citealt{Sathyaprakash+Dhurandhar+1991})
% 在受限2-PN框架下，频域中的两个正交极化波形$\tilde{h}+(f)$和$\tilde{h}\times(f)$可表示为（\citealt{Sathyaprakash+Dhurandhar+1991}）
\begin{align}
    \begin{split}
        \tilde{h}_+(f) = \frac{A_f}{r} \frac{(1+\cos^2\iota)}{2} f^{-7/6} \exp{[-i\Psi(f)]} \\
        \tilde{h}_\times(f) = \frac{A_f}{r} \cos\iota f^{-7/6} \exp{[-i\Psi(f)+\pi/2]} 
    \end{split}
\end{align}
where $r$ is the source distance, $\iota$ is the orbital inclination angle, and $A_f$ is an amplitude coefficient related to the chirp mass of the binary system. The frequency-domain phase $\Psi(f)$ is determined by the intrinsic parameters of the binary system and the time of coalescence, and its expansion is
% 其中，$r$为源距离，$\iota$为轨道倾角,$A_f$为与双星系统啁啾质量相关的振幅系数。频域相位$\Psi(f)$由双星系统的固有参数和合并时刻决定，其展开形式为
\begin{equation}
    \Psi =2\pi f t_c - \psi_c -\pi/4 +\sum_{j=0}^4 a_j(\frac{f}{f_*})^{(-5+j)/3}
\end{equation}
where $t_c$ is the coalescence time, $\psi_c$ is the coalescence phase, and $f_*$ is the prescribed low-frequency cutoff. The phase expansion coefficients $\alpha_j$ are determined by the chirp time. Under the constraint that the component masses satisfy $m_1 > m_2$, the chirp time can be written as
% 式中，$t_c$为合并时间，$\psi_c$为合并相位，$f_*$为设定的低频截至频率。相位展开系数$\alpha_j$由啁啾时间决定。在约束双星质量满足$m_1>m_2$的情况下，啁啾时间可写为
\begin{align}
    \tau_0 &= \frac{5}{256} f_*^{-1} \left( \frac{GM}{c^3} \pi f_* \right)^{-5/3} \eta^{-1}  \label{tau_0}\\
    \tau_1 &= \frac{5}{192\pi} f_*^{-1} \left( \frac{GM}{c^3} \pi f_* \right)^{-1} \eta^{-1} \left( \frac{743}{336} + \frac{11}{4}\eta \right) \\
    \tau_{1.5} &= \frac{1}{8} f_*^{-1} \left( \frac{GM}{c^3} \pi f_* \right)^{-2/3} \eta^{-1}  \label{tau15}\\
    \tau_2 &= \frac{5}{128\pi} f_*^{-1} \left( \frac{GM}{c^3} \pi f_* \right)^{-1/3} \eta^{-1}
            \left( \frac{3058673}{1016064} + \frac{5429}{1008}\eta + \frac{617}{144}\eta^2 \right)
\end{align}
$M$, $\mu$, and $\eta$ are the total mass, reduced mass, and symmetric mass ratio of the binary system, respectively, defined as
% $M$,$\mu$和$\eta$分别为双星系统的总质量、约化质量和对称质量比，其定义为
\begin{equation}
    M = (m_1+m_2),\quad \quad \mu = \frac{m_1m_2}{M},\quad \quad\eta = \frac{\mu}{M}
    \label{M}
\end{equation}
Thus, the phase expansion coefficients $\alpha_j$ of $\Psi(f)$ can be written as
% 由此，$\Psi(f)$的相位展开系数$\alpha_j$可写为
\begin{equation}
\begin{split}
    \alpha_0 &= 2\pi f_* \frac{3\tau_0}{5}, \quad \alpha_1 = 0, \quad \alpha_2 = 2\pi f_* \tau_1 \\
    \alpha_3 &= -2\pi f_* \frac{3\tau_{1.5}}{2}, \quad \alpha_4 = 2\pi f_* 3\tau_2
\end{split}
\label{lab:}
\end{equation}

To construct a fully coherent search statistic for a multidetector network and avoid nonlinear searches over parameters such as $r$, $\iota$, $\psi$, and $\phi$, we absorb these parameters into four linearly independent amplitude parameters, $A_k$ $(k=1,2,3,4)$. This reduces the dimensionality of the search space. Consequently, the response of detector $J$ can be written as a linear combination of four reference waveform sequences:
% 为了在多探测器网络中构建全相干搜索统计量，并避免对$r$、$\iota$、$\psi$和$\phi$等参数进行非线性搜索，我们将这些参数吸收到四个线性独立的振幅参数$A_k$（$k=1,2,3,4$）中。这降低了搜索空间的维度。因此，第$I$个探测器的响应可以写成四个参考波形序列的线性组合：
\begin{equation}
    h^J(t) = \sum_{k=1}^4 A_k h_k^J(t)
    \label{signal_1}
\end{equation}
where,
% 其中，
\begin{equation}
    \begin{split}
        A_1 = \frac{1}{r}(\beta\cos\phi_c\cos2\psi - \cos\iota\sin\phi_c\sin2\psi)\\
        A_2 = \frac{1}{r}(\beta\cos\phi_c\sin2\psi + \cos\iota\sin\phi_c\cos2\psi)\\
        A_3 = -\frac{1}{r}(\beta\sin\phi_c\cos2\psi + \cos\iota\cos\phi_c\sin2\psi)\\
        A_4 = -\frac{1}{r}(\beta\sin\phi_c\sin2\psi - \cos\iota\cos\phi_c\cos2\psi)
    \end{split}
    \label{signal_2}
\end{equation}
where $\beta = (1+\cos^2\iota)/2$, and the reference templates $h^J_k$ are defined as
% 式中$\beta = (1+\cos^2\iota)/2$,基准模板$h^I_k$定义为
\begin{equation}
    \begin{split}
        h_1^J(t) = U_+^Jh_c(t - \Delta^J)   \quad\quad  h_2^J(t) = U_\times^Jh_c(t - \Delta^J)\\
        h_3^J(t) = U_+^Jh_s(t - \Delta^J)   \quad\quad  h_4^J(t) = U_\times^Jh_s(t - \Delta^J)
    \end{split}
    \label{signal_3}
\end{equation}
where $\tilde{h}_c(f)$ denotes the normalized frequency-domain waveform when the coalescence phase $\phi_c = 0$, with $\tilde{h}_c(f) = A_f f^{-7/6} \exp[-i\Psi(f)|{\phi_c = 0}]$. $\tilde{h}_s(f)$ denotes its phase-shifted counterpart and satisfies the relation $\tilde{h}_s(f) = -i\tilde{h}_c(f)$.
% 其中，$\tilde{h}_c(f)$表示合并相位$\phi_c =0$时的标准化频域波形，有$\tilde{h}_c(f) = A_f f^{-7/6} \exp[-i\Psi(f)|_{\phi_c = 0}]$。$\tilde{h}_s(f)$表示其相移形式，并满足关系$\tilde{h}_s(f) =  -i\tilde{h}_c(f)$。

In addition to the restricted 2-PN model, we use the IMRPhenomD waveform as an injected signal (\citealt{Husa+etal+2016}) to test the robustness of the algorithm
under different waveform-model assumptions. In GW searches, systematic errors may arise from truncated approximation orders or missing physical effects, leading to mismatches between the template and the true signal (\citealt{Abbott+etal+2017}). We therefore use the relatively simple restricted 2-PN template to recover the more complete IMRPhenomD signal, thereby evaluating the impact of template-signal mismatch on algorithmic performance, parameter bias, and classification stability. Thus, this work evaluates algorithmic performance under model-matched conditions and assesses the usability and generalizability of the algorithm under nonideal template conditions (\citealt{Dhani+etal+2025}). For the two signal models, restricted 2-PN and IMRPhenomD, we adopt a 70:30 ratio. This allocation enables a detailed assessment of algorithmic performance under model-matched conditions while retaining sufficient samples to examine generalizability.
% 除受限2-PN模型外，我们还使用IMRPhenomD波形作为注入信号（\citealt{Husa+etal+2016}），以检验算法在不同波形模型假设下的稳健性。在引力波搜索中，系统误差可能源于近似阶数的截断或物理效应的缺失，从而导致模板与真实信号之间的不匹配（\citealt{Abbott+etal+2017}）。因此，我们使用相对简单的受限2-PN模板来恢复更完整的IMRPhenomD信号，从而评估模板-信号不匹配对算法性能、参数偏差和分类稳定性的影响。因此，本文既评估模型匹配条件下的算法性能，也评估算法在非理想模板条件下的可用性和泛化能力（\citealt{Dhani+etal+2025}）。对于受限2-PN和IMRPhenomD这两种信号模型，我们采用70:30的比例。这种分配方式能够充分评估模型匹配条件下的算法性能，同时保留足够样本以考察算法的泛化能力。

%%======================模型无关的双象透镜=================================
\subsection{Model-independent image parameterization}
\begin{figure}[htbp]
    \centering
    \includegraphics[width=12cm, angle=0]{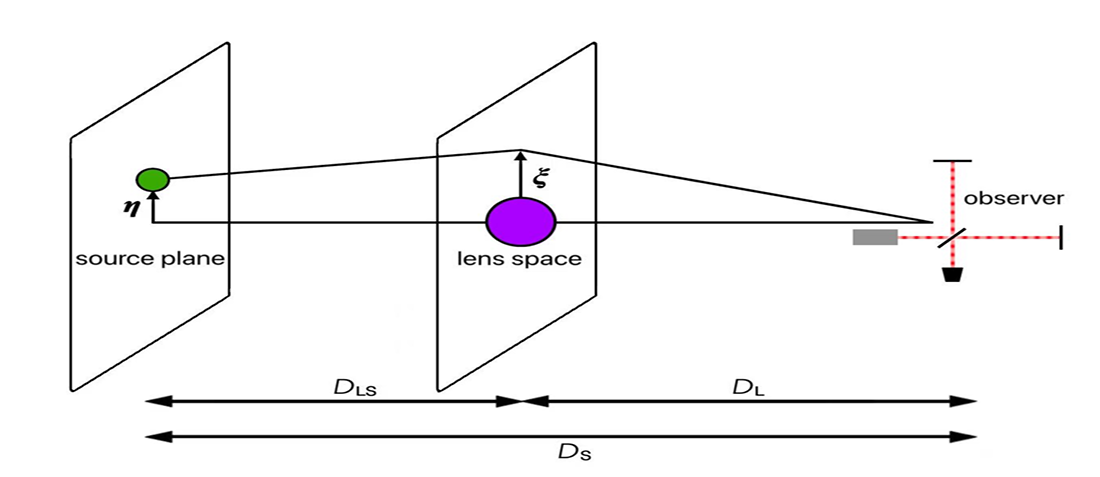}
    \caption{Schematic diagram of the gravitational lens system. The system consists of a GW source on the source plane, a lensing object on the lens plane, and the observer. Where $D_L$, $D_{LS}$, and $D_S$ denote the angular diameter distances from the observer to the lens, from the lens to the source, and from the observer to the source, respectively.  $\bm{\xi}$ denotes the impact parameter on the lens plane \citep{Qin+etal+2025}.}
    \label{fig:Fig1}
\end{figure}

As shown in Fig.~\ref{fig:Fig1}, when a GW passes near a compact object or galaxy close to the line of sight, its trajectory is deflected by the gravitational potential of the lens, producing a gravitational-lensing effect (\citealt{Schneider+etal+1992}).
% 如图~\ref{fig:Fig1}所示，当引力波（GW）在传播过程中经过靠近视线方向的致密天体或星系附近时，其轨迹会受到透镜引力势的偏折，从而产生引力透镜效应（\citealt{Schneider+etal+1992}）。
For a lensed GW, its frequency-domain waveform can be expressed as
% 对于透镜化引力波信号，其频域波形可表示为
\begin{equation}
    \tilde{h}_L(f)=F(f)\tilde{h}(f),
\end{equation}
where $\tilde{h}(f)$ denotes the unlensed GW signal, $\tilde{h}_L(f)$ denotes the lensed GW signal, and $F(f)$ is the lensing amplification factor. Therefore, when constructing the dataset, we first generate the unlensed GW signal and then modulate it by the lensing amplification factor to obtain the lensed GW data.
% 其中，$\tilde{h}(f)$ 表示为未透镜化引力波，$\tilde{h}_L(f)$表示透镜化引力波，$F(f)$为透镜放大因子。因此，在构造数据集时，只需先生成未透镜化引力波，再通过透镜放大因子对其进行调制，即可得到透镜化引力波数据。

Under the thin-lens approximation, the lensing system is described by three components: the source plane, the lens plane, and the observer. For an axisymmetric lens, the source-position vector $\bm{\eta}$ and impact-parameter vector $\bm{\xi}$ reduce to the scalars $\eta$ and $\xi$, respectively. The dimensionless source-position parameter is then defined as
% 在薄透镜近似下，透镜系统由三个组成部分描述：源平面、透镜平面和观测者。对于轴对称透镜，源位置矢量$\bm{\eta}$和冲量参数矢量$\bm{\xi}$可分别简化为标量$\eta$和$\xi$。随后，无量纲源位置参数定义为：
\begin{equation}
    y \equiv \frac{\eta}{\xi_0 D_S / D_L},
\end{equation}
where $\xi_0$ is the characteristic scale on the lens plane. For the singular isothermal sphere (SIS) lens model, the Einstein radius can be written, following ~\citet{Ali+etal+2023}, as
% 其中，$\xi_0$ 为透镜平面上的特征尺度。对于奇异等温球（SIS）透镜模型，Einstein 半径参考~\citealt{Ali+etal+2023}可写为
\begin{equation}
    \xi_0 = 4\pi(\sigma_v/c)^2 \dfrac{D_L D_{LS}}{D_S},
\end{equation}
where $\sigma_v$ is the velocity dispersion of the lensing object, and $D_L$, $D_{LS}$, and $D_S$ denote the angular diameter distances from the observer to the lens, from the lens to the source, and from the observer to the source, respectively.
% 其中，$\sigma_v$ 为透镜天体的速度弥散，$D_L$、$D_{LS}$ 和 $D_S$ 分别表示观测者到透镜、透镜到源以及观测者到源的角直径距离。

We adopt the SIS lens model and study lensed GWs under the geometric-optics approximation. This approximation is valid when the GW wavelength is much smaller than the characteristic scale of the lens \citep{JitSingh+etal+2018,Haris+etal+2018}. Under this approximation, a lensed GW can be treated as the coherent superposition of several discrete images. For the SIS model, two images are formed when the source lies within the Einstein radius ($y<1$), whereas only one image is formed when $y>1$. Accordingly, under the geometric-optics approximation, the frequency-domain amplification factor can be written as follows (\citealt{Ali+etal+2023}):
% 本文采用 SIS 透镜模型，并在几何光学近似下研究透镜化引力波。当引力波波长远小于透镜的特征尺度时，该近似是有效的（\citealt{JitSingh+etal+2018,Haris+etal+2018}）。在这一近似下，透镜化引力波可以被视为若干离散像的相干叠加。对于SIS模型，当源位于爱因斯坦半径以内（$y<1$）时，会成两个像；而当$y>1$时，只形成一个像。因此，在几何光学近似下，频域放大因子可写为（\citealt{Ali+etal+2023}）：
\begin{equation}
    F(f)=
    \begin{cases}
        |\mu_+|^{1/2}-i|\mu_-|^{1/2}e^{2\pi i f \Delta t_d}, & y<1,\\[6pt]
        |\mu_+|^{1/2}, & y>1,
    \end{cases}
\end{equation}
where $\mu_\pm$ denote the magnifications of the two images, respectively, and $\Delta t_d$ is the time delay between the two images. For the SIS model, under the above notation convention, we have
% 其中，$\mu_\pm$分别表示两幅像的放大率，$\Delta t_d$为双象之间的时间延迟。对于 SIS模型，在上述记号约定下，有
\begin{equation}
    \mu_{\pm}=\pm 1+\frac{1}{y}
    \label{mu}
\end{equation}
\begin{equation}
    \Delta t_d = 8 M_L y
    \label{td}
\end{equation}
where $M_L$ denotes the effective lens mass within the Einstein radius.
% 其中，$M_L$ 表示 Einstein 半径内的有效透镜质量。

To reduce the dependence of the search and parameter estimation on a specific lens model, we do not directly use the lens parameters $(M_L,y)$ to describe the
lensing effect. Instead, we adopt two model-independent image parameters: the flux ratio $I$ and the time delay $\Delta t_d$. The flux ratio is defined as
% 为了降低搜索和参数估计对特定透镜模型的依赖，我们不直接使用透镜参数$(M_L,y)$来描述透镜效应。相反，我们采用两个模型无关的像参数：流量比$I$和时间延迟$\Delta t_d$。流量比定义为：
\begin{equation}
    I \equiv \frac{|\mu_-|}{|\mu_+|}
    \label{I}
\end{equation}

The above equation shows that, in the double-image regime under the geometric-optics approximation, the interference modulation of a lensed GW is primarily governed by the image flux ratio $I$ and the time delay $\Delta t_d$. Therefore, using $(I,\Delta t_d)$ as dataset labels and parameter-estimation variables directly characterizes the observable modulation features induced by lensing while reducing dependence on specific lens models. Based on this parameterization, the following sections construct a lensed GW dataset in the image-parameter space and investigate signal searches and parameter estimation using particle swarm optimization.
% 上述方程表明，在几何光学近似下的双像情形中，透镜化引力波的干涉调制主要由像流量比$I$和时间延迟$\Delta t_d$控制。因此，将$(I,\Deltat_d)$作为数据集标签和参数估计变量，可以直接表征由透镜效应引起的可观测调制特征，同时降低对特定透镜模型的依赖。基于这一参数化，后续章节将在像参数空间中构建透镜化引力波数据集，并利用粒子群优化方法研究信号搜索和参数估计。

%%===============================================全相干网络===========================================
\subsection{Fully coherent all-sky search}
In what follows, the strain time series $s^J(t)$ recorded by the $J$-th detector can be expressed as $s^J(t) = h^J(t) + n^J(t)$, where $h^J(t)$ is the GW response at that detector and $n^J(t)$ is the detector noise (\citealt{Jaranowski+Krolak+2012}).
% 在下文中，第$I$个探测器记录的应变时间序列$s^I(t)$可表示为$s^I(t)=h^I(t)+n^I(t)$，其中$h^I(t)$为该探测器中的引力波响应，$n^I(t)$为探测器噪声（\citealt{Jaranowski+Krolak+2012}）。

In the fully coherent detection network, the log-likelihood ratio (LLR) for the $J$-th detector $\ln \lambda^J$ can be written as (\citealt{Helstrom+1995}):
% 在全相干探测网络中，第$I$个探测器的对数似然比(LLR)(~\citealt{Helstrom+1995})可写为：
\begin{equation}
    \ln \lambda^J = \langle s^J|h^J\rangle - \frac{1}{2}\langle h^J |h^J\rangle,
\end{equation}
where inner product is defined as
\begin{equation}
    \langle s|h \rangle = 4 \mathrm{Re} \int_{0}^{\infty} df \dfrac{\tilde{s}^{*}(f)\tilde{h}(f)}{S_n(f)},
\end{equation}
and $s^J$ denotes the data received by the detector, and $h^J$ is the GW template, $S_n(f)$ is the detector noise power spectral density（PSD）. To simplify the simulation, we assign the same PSD, generated using \texttt{pycbc.psd.aLIGOZeroDetHighPower} \citep{LALSuite+2020}, to all four detectors. Assuming that the noise is independent across detectors, an $N$ detector network log-likelihood can be written as
% 式中$S_n(f)$为探测器的噪声功率谱密度。为简化模拟，本实验中四个探测器均采用由\textbf{pycbc.psd.aLIGOZeroDetHighPower}(~\citealt{LALSuite+2020})生成的相同PSD。若假设不同探测器中的噪声是独立的，则这个N探测器网络的对数似然可写为
\begin{equation}
    \ln \lambda^N = \sum_{J\in\{H_1,L_1,V_1,K_1\}}[\langle s^J|h^J \rangle - \frac{1}{2} \langle h^J|h^J \rangle]
\end{equation}
Substituting Eqs.\eqref{signal_1}, \eqref{signal_2}, and \eqref{signal_3} into the above equation yields
% 将公式~\ref{signal_1}、~\ref{signal_2}、~\ref{signal_3}带入上式，可得
\begin{equation}
    \ln \lambda^N = A^TH - \frac{1}{2}A^TMA
\end{equation}
where $\boldsymbol{A} =(A_1,A_2,A_3,A_4)^T$ and $\boldsymbol{H} =(H_1,H_2,H_3,H_4)^T$. For $a = 1,\dots,4$, we have $H_a = \sum_{J\in{H_1,L_1,V_1,K_1}}\langle s^J|h_a^J \rangle$. The elements of the network response matrix $M$ can be obtained by summing the inner products of the reference waveforms across all detectors:
% 其中，$\boldsymbol{A} =(A_1,A_2,A_3,A_4)^T$,$\boldsymbol{H} = (H_1,H_2,H_3,H_4)^T$。对于$a = 1,\dots,4$有
% $H_a = \sum_{I\in\{H_1,L_1,V_1,K_1\}}\langle x^I|h_a^I \rangle$ 。网络响应矩阵$M$的元素可由所有探测器上的基准波形内积累加得到，即
\begin{equation}
    M_{ab} = \sum_{J\in\{H_1,L_1,V_1,K_1\}} \langle h_a^J|h_b^J \rangle \quad  a,b \in\{1,2,3,4\}
\end{equation}
It follows that, for a given data sequence, the log-likelihood ratio is a function of the linear parameters $\boldsymbol{A}$ and the nonlinear parameters set $\Theta = {\tau_0,\tau_{1.5},I,t_d,\alpha,\delta}$. The fully coherent search statistic is defined as the likelihood ratio statistic obtained by globally maximizing over $\boldsymbol{A}$, $t_c$, and $\Theta$. It can be written as
% 由此可知，对于给定的数据序列，对数似然比是线性参数$\boldsymbol{A}$与非线性参数集合$\Theta = \{\tau_0,\tau_{1.5},I,t_d,\alpha,\delta\}$的函数。全相干搜索统计量定义为对参数$\boldsymbol{A}$,$t_c$和$\Theta$做全局极大化后得到的似然比统计量。可写为
\begin{equation}
    \rho_{coh}^2 = 2 \max_{A,t_c,\Theta} \ln \lambda^N
\end{equation}
The maximization of the log-likelihood ratio can be performed sequentially as follows:
\begin{equation}
    \begin{split}
        \rho_{coh}^2 = \max_{\Theta} \gamma^2(\Theta) \\
        \gamma^2(\Theta) = 2 \max_A \ln \lambda^N
    \end{split}
\end{equation}
First, analytical maximization over $\boldsymbol{A}$ gives $\boldsymbol{\hat{A}}=M^{-1}\boldsymbol{H}$ and hence $\gamma^2(\Theta)=\boldsymbol{H}^{T}
M^{-1}\boldsymbol{H}$. For fixed $\Theta$, the coalescence time $t_c$ can be efficiently scanned and maximized using the fast Fourier transform, yielding $
\Gamma^2(\Theta)=\max_{t_c}\gamma^2(\Theta)$ (\citealt{Bose+etal+2011}). The key step is therefore to identify the parameter set $\hat{\Theta}$ that maximizes the statistic. In this study, we use PSO to solve this optimization problem. Table~\ref{data parameter range} lists the initial parameter ranges used in this work.
% 首先，对$\boldsymbol{A}$进行解析极大化可得$\boldsymbol{\hat{A}}=M^{-1}\boldsymbol{H}$，因此$\gamma^2(\Theta)=\boldsymbol{H}^{T}M^{-1}\boldsymbol{H}$。在固定$\Theta$的情况下，并合时间$t_c$可通过快速傅里叶变换进行高效扫描和极大化，从而得到$\Gamma^2(\Theta)=\max_{t_c}\gamma^2(\Theta)$（\citealt{Bose+etal+2011}）。因此关键步骤是确定使统计量最大的参数集合$\hat{\Theta}$。在本研究中，我们使用PSO求解这一优化问题。表~\ref{data parameter range}列出了本文使用的初始参数范围。
\begin{table}[htbp]
\centering
\caption{Parameter descriptions for the binary system model}
\begin{tabular}{lllll}
\hline
\textbf{Parameter} & \textbf{Description} & \textbf{Range} & \textbf{Unit} & \textbf{Distribution} \\
\hline
$m_1$       & Primary black hole mass    & $(20,\,50)$           & $M_\odot$  & Uniform \\
$m_2$       & Secondary black hole mass   & $(20,\,50)$           & $M_\odot$  & Uniform \\
$r$         & Luminosity distance to the source & $(200,\,5000)$  & Mpc        & Uniform    \\
$\iota$       & Angle of inclination of the binary orbit                             & $(0,\,\pi)$      & rad  & Sine \\
$\phi_c$       &  Phase at the moment of coalescence        & $(0,\,2\pi)$            & rad        & uniform   \\
$\alpha$    & Right Ascension       & $(0,\,2\pi)$         & rad        & Uniform    \\
$\delta$    & Declination        & $(-\pi/2,\,\pi/2)$   & rad        & Cosine     \\
$\psi_s$    & Polarization angle of the gravitational wave   & $(0,\,\pi)$          & rad        & Uniform    \\
$t_c$      & Time of coalescence       & $(1.2,\,1.8)$          & s        & uniform       \\
$y$ & Impact parameter of the lensing system             & $(0,\,1)$         & $-$        & $p(y)\propto \sqrt{y}$    \\
$M_l$    & Mass of the lens                            & $(10^3,\,10^5)$            & $M_\odot$          & LogUniform    \\
\hline
\end{tabular}
\label{data parameter range}
\end{table}

%%=================================================粒子群算法============================================
\section{Particle Swarm Optimization}
\label{sect:PSO}
PSO is a stochastic optimization method based on collective behavior and is well suited to high-dimensional, nonlinear, and non-
differentiable parameter-search problems (\citealt{Kennedy+Eberhart+1995,Shi+2004}). In PSO, each particle represents a candidate solution in the parameter space.
During the search, particles iteratively update their velocities and positions by combining their personal best solutions with the global best solution found by the swarm, thereby gradually approaching an optimum.
% 粒子群优化（PSO）是一种基于群体行为的随机优化方法，适用于高维、非线性和不可微的参数搜索问题（\citealt{Kennedy+Eberhart+1995,Shi+2004}）。在PSO中，每个粒子代表参数空间中的一个候选解。在搜索过程中，粒子通过结合自身历史最优解和群体已找到的全局最优解，迭代更新其速度和位置，从而逐渐接近一个最优解。

Let the position and velocity vectors of the ($i$)th particle at iteration ($t$) be $(\boldsymbol{x}_i^t)$ and $(\boldsymbol{v}_i^t)$, respectively. The velocity and
position are then updated as follows:
% 设第(i)个粒子在第(t)次迭代时的位置矢量和速度矢量分别为(\boldsymbol{x}_i^t)和(\boldsymbol{v}_i^t)。其速度和位置按如下方式更新：
\begin{equation}
    \boldsymbol{v}_i^{t+1}
    =
    w^t \boldsymbol{v}_i^t
    + c_1 r_1 \left(\boldsymbol{p}_i - \boldsymbol{x}_i^t\right)
    + c_2 r_2 \left(\boldsymbol{g} - \boldsymbol{x}_i^t\right),
    \label{PSO1}
\end{equation}
\begin{equation}
    \boldsymbol{x}_i^{t+1} = \boldsymbol{x}_i^t + \boldsymbol{v}_i^{t+1}.
    \label{PSO2}
\end{equation}
Here, $(\boldsymbol{p}_i)$ denotes the personal best position attained by particle ($i$) over its iteration history, and $(\boldsymbol{g})$ denotes the global best position of the swarm. The variables $(r_1)$ and $(r_2)$ are independent random numbers uniformly sampled from the interval ([0,1]). The coefficients $(c_1)$ and $(c_2)$ are the cognitive and social learning factors, respectively, and control the extent to which a particle moves toward its personal best position and the global best position. In Eq.~\eqref{PSO1}, the first term is the inertia term, which preserves the particle's previous direction of motion. The second and third terms correspond to the cognitive and social components, respectively.
%  其中，(\boldsymbol{p}_i)表示粒子(i)在迭代历史中达到的个体最优位置，(\boldsymbol{g})表示粒子群的全局最优位置。变量(r_1)和(r_2)是从区间([0,1])中均匀采样的相互独立随机数。系数(c_1)和(c_2)分别为认知学习因子和社会学习因子，用于控制粒子向其个体最优位置和全局最优位置移动的程度。在式~\eqref{PSO1}中，第一项为惯性项，用于保留粒子先前的运动方向；第二项和第三项分别对应认知分量和社会分量。

As indicated by Eqs.~\eqref{PSO1} and \eqref{PSO2}, the PSO search process involves a trade-off between global exploration and local exploitation. The inertia weight $(w^t)$ plays a key role in controlling this balance (\citealt{Shi+Eberhart+1998}). A larger inertia weight helps maintain the continuity of particle motion and thereby enhances global exploration of the parameter space. In contrast, a smaller inertia weight facilitates finer local exploitation within high-fitness regions \citep{Bansal+etal+2011}. To balance early-stage exploration with late-stage convergence stability, we adopt a linearly decreasing inertia-weight strategy:
% 如式~\eqref{PSO1}和\eqref{PSO2}所示，PSO搜索过程涉及全局探索与局部开发之间的权衡。惯性权重(w^t)在控制这一平衡中起关键作用（\citealt{Shi+Eberhart+1998}）。较大的惯性权重有助于保持粒子运动的连续性，从而增强参数空间中的全局探索能力。相比之下，较小的惯性权重有利于粒子在高适应度区域内进行更精细的局部开发（\citealt{Bansal+etal+2011}）。为了平衡早期阶段的探索能力与后期阶段的收敛稳定性，我们采用线性递减的惯性权重策略：
\begin{equation}
    w^t = w_{\mathrm{start}}
    - \left(w_{\mathrm{start}} - w_{\mathrm{end}}\right)\frac{t}{T_{\max}},
\end{equation}
where $(w_{\mathrm{start}})$ and $(w_{\mathrm{end}})$ denote the initial and final inertia weights, respectively, and $(T_{\max})$ is the maximum number of iterations. This strategy gives the swarm stronger global exploration capability during the early iterations and gradually improves local convergence in later iterations.
% 其中，(w_{\mathrm{start}})和(w_{\mathrm{end}})分别表示初始和最终惯性权重，(T_{\max})为最大迭代次数。该策略使粒子群在早期迭代中具有更强的全局探索能力，并在后期迭代中逐渐改善局部收敛性能。

In our parameter-estimation task, the position vector of each particle represents a candidate parameter set, and its fitness is determined by the network coherent
statistic computed for that parameter combination. The fitness function is defined as
% 在我们的参数估计任务中，每个粒子的位置矢量表示一个候选参数集合，其适应度由该参数组合下计算得到的网络相干统计量决定。适应度函数定义为：
\begin{equation}
    \mathrm{Fitness}(\boldsymbol{x}_i^t)=\rho_{\mathrm{coh}}^2(\boldsymbol{x}_i^t),
\end{equation}
where $\rho_{\mathrm{coh}}^2$ denotes the network coherent statistic. The objective of PSO is to maximize this fitness function over the prescribed parameter space, thereby obtaining the best-fit parameter estimates. In simulated-injection experiments, the coherent statistic corresponding to the true parameters is used only as a reference benchmark for evaluating algorithmic convergence, not as a target value known in advance during optimization (\citet{Wang+Mohanty+2010}).
% 其中，(\rho{\mathrm{coh}}^2)表示网络相干统计量。PSO的目标是在给定参数空间内最大化该适应度函数，从而获得最佳拟合参数估计值。在模拟注入实验中，真实参数对应的相干统计量仅作为评估算法收敛性的参考基准，而不是优化过程中预先已知的目标值（\citealt{Wang+Mohanty+2010}）。

All search parameters are linearly normalized to the interval ([0,1]) before being passed to PSO. To prevent excessively large particle displacements in a single
update, we impose an upper bound $v_{\max}$ on the particle velocity. We adopt the standard settings $c_1=c_2=2$, $w_{\mathrm{start}}=0.9$, $w_{\mathrm{end}}=0.4$,
and $v_{\max}=0.5$ (\citealt{Poli+etal+2007}). Based on repeated experiments, we set the swarm size to $N_{\mathrm{pop}}=50$ and the maximum number of iterations to $N_{\mathrm{iter}}=1000$.
% 所有搜索参数在输入PSO之前都被线性归一化到区间([0,1])。为防止粒子在单次更新中产生过大的位移，我们对粒子速度施加上限v_{\max}。我们采用标准设置c_1=c_2=2、w_{\mathrm{start}}=0.9、w_{\mathrm{end}}=0.4和v_{\max}=0.5（\citealt{Poli+etal+2007}）。基于多次实验，我们将粒子群规模设为N_{\mathrm{pop}}=50，最大迭代次数设为N_{\mathrm{iter}}=1000。

\begin{figure}[htbp]
  \centering
  \includegraphics[width=\textwidth]{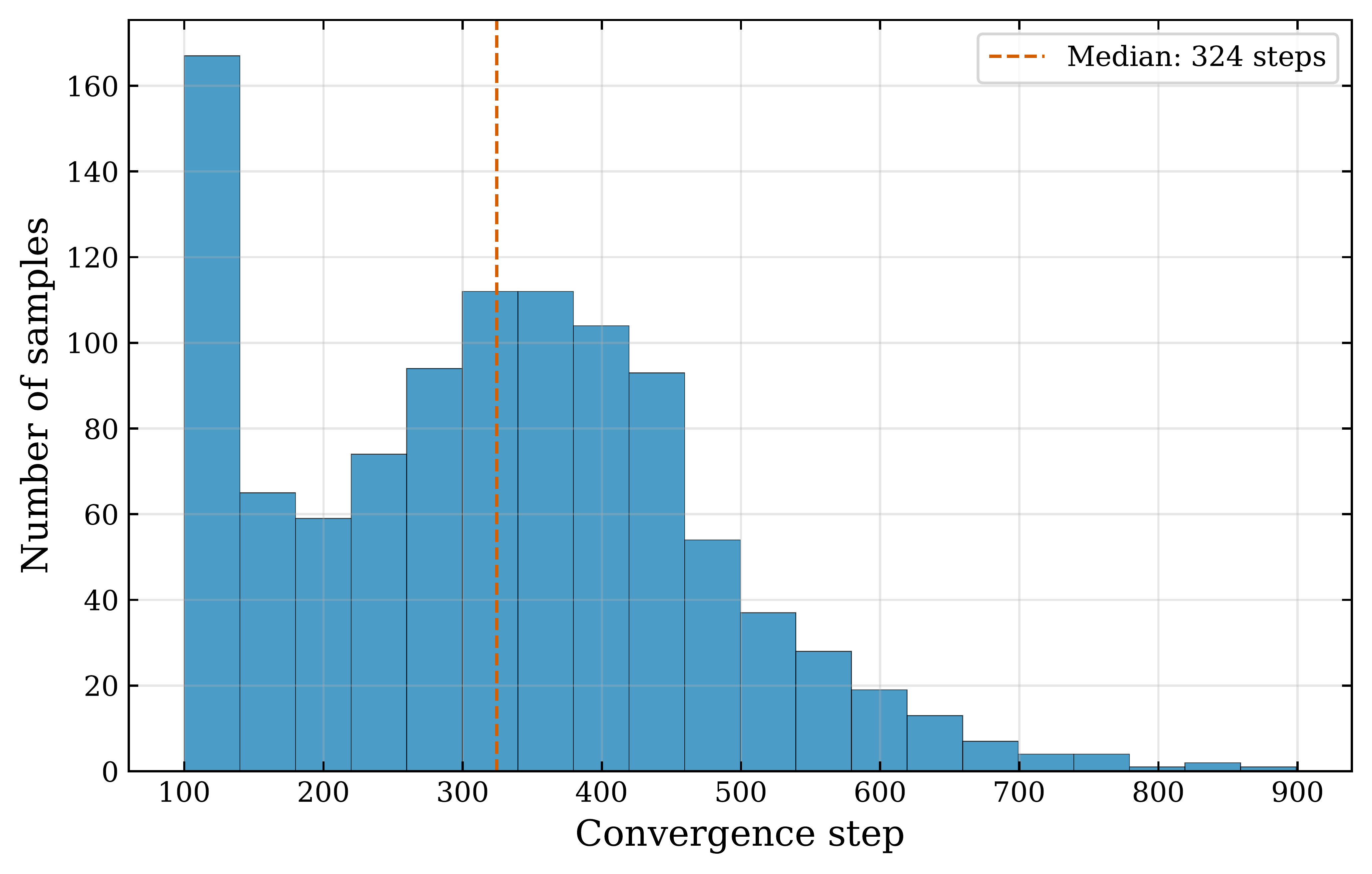}
  \caption{The histogram of convergence step count. The mean is 320 steps,the median is 324 steps,the maximum is 899 steps. }
  \label{fig:steps}
\end{figure}

Figure~\ref{fig:steps} shows the distribution of the number of iterations required for convergence across all samples. The mean and median numbers of convergence steps are 320 and 324, respectively, and the maximum is 899; all samples converge within the 1,000-iteration budget, with about 89\% of them converging within 500 steps. The current settings ($N_{\mathrm{pop}}=50$ and $N_{\mathrm{iter}}=1000$) are therefore sufficient to guarantee convergence for all samples. For more complex waveforms, such as those involving spin precession, the required population size and iteration count are expected to increase with the dimensionality of the parameter space. For instance, in a five-dimensional search for gravitational waves from binaries with eccentricity and spins, Pal and Nayak (2024) used 5,000 particles and at most 15 iterations, amounting to about 75,000 template evaluations (\citealt{Pal+Nayak+2024}); for the more complex ten-dimensional search of extreme-mass-ratio inspiral signals, Zou et al. adopted 40 particles, 10,000 iterations, and 6 independent runs (\citealt{Zou+Mohanty+2024}). These examples demonstrate that, for more complex waveforms and higher-dimensional parameter estimation, increasing the swarm size and the number of iterations according to the requirements of the specific problem is necessary.
% 我们绘制了全部数据收敛所需迭代次数的直方图，如图~\ref{fig:steps}所示。其中，平均收敛步数为320步，中位数为324步，最大迭代步数为899步，均在1000步之内，且约有89\%的数据在500步之内便达到收敛，因此当前$N_{\mathrm{pop}}$以及$N_{\mathrm{iter}}$的设置能够保证所有样本均达到收敛状态。对于包含自旋进动或更复杂的波形，随着参数维度的升高，所需的$N_{\mathrm{pop}}$以及$N_{\mathrm{iter}}$也应有所升高。Pal和Nayak在2024年对含偏心率和自旋的五维引力波搜索中使用了5000个粒子、最多15步迭代，总匹配次数约为75000次(\citealt{Pal+Nayak+2024})；Zou等人在十维更复杂的extreme-mass-ratio inspiral signals的搜索中使用了40个粒子、10000次迭代、6次独立运行进行研究(\citealt{Zou+Mohanty+2024})。因此，在更复杂的波形以及更高维的参数估计中，根据实际问题需求提高粒子种群数量以及迭代次数是有必要的。

In the following sections, we evaluate the performance of PSO in lensed GW searches, parameter point estimation, and classification using simulated datasets. We further
examine its impact on posterior estimation and computational efficiency when combined with nested sampling . The dataset contains 1050 samples: 350 lensed GW samples, 350 unlensed GW samples, and 350 noise-only samples. For the signal samples, we consider restricted 2-PN signals under model-matched conditions and IMRPhenomD signals under template-signal mismatch conditions. Throughout this work, we use the lensed template to match all 1050 samples and run 12 independent PSO searches in parallel for each sample to examine the stability and generalizability of PSO under different waveform models.
% 在以下各节中，我们基于模拟数据集评估PSO在透镜化引力波搜索、参数恢复和分类任务中的性能。我们进一步考察PSO与嵌套采样结合时对后验估计和计算效率的影响。该数据集包含1050个样本：350个透镜化引力波样本、350个未透镜化引力波样本和350个纯噪声样本。对于信号样本，我们同时考虑模型匹配条件下的受限2-PN信号和模板-信号不匹配条件下的IMRPhenomD信号。在整个研究中，我们使用透镜化模板匹配全部1050个样本，并对每个样本并行运行12次独立PSO搜索，以检验PSO在不同波形模型下的稳定性和泛化能力。

The search ranges of the six PSO parameters are listed in Table~\ref{tab:parameter search range}. To evaluate the global search capability of PSO as broadly as
possible, we set the search range of each parameter to its full allowed extent.
% 六个PSO参数的搜索范围列于表~\ref{tab:parameter search range}。为了尽可能广泛地评估PSO的全局搜索能力，我们将每个参数的搜索范围设置为其允许的完整范围。
\begin{table}[htbp]
\centering
\caption{Parameter search range}
\begin{tabular}{lllll}
\hline
\textbf{Parameter} & \textbf{Description} & \textbf{Range} & \textbf{Unit} & \textbf{Distribution} \\
\hline
$\tau_0$       & Chirping Time      & $(0,\,2)$               & s       & LogUniform   \\
$\tau_{1.5}$   & Chirping Time      & $(0,\,2)$               & sec      & Uniform      \\
$\alpha$       & Right Ascension    & $(0,\,2\pi)$            & rad      & Uniform      \\
$\delta$       & Declination        & $(-\pi/2,\,\pi/2)$      & rad      & Cosine       \\
$I$            & Flux ratio         & $(0,\,1)$               & $-$      &$p(I)\propto 1/y$      \\
$t_d$          & Time delay         & $(0,\,2)$       & sec      & $p(t_d) \propto y\cdot M_l$    \\
\hline
\end{tabular}
\label{tab:parameter search range}
\end{table}

%%============================参数估计===============================
\section{Parameter point estimation }
\label{sect:Parameter recovery}
\begin{figure}[htbp]
    \centering
    \includegraphics[width=\textwidth]{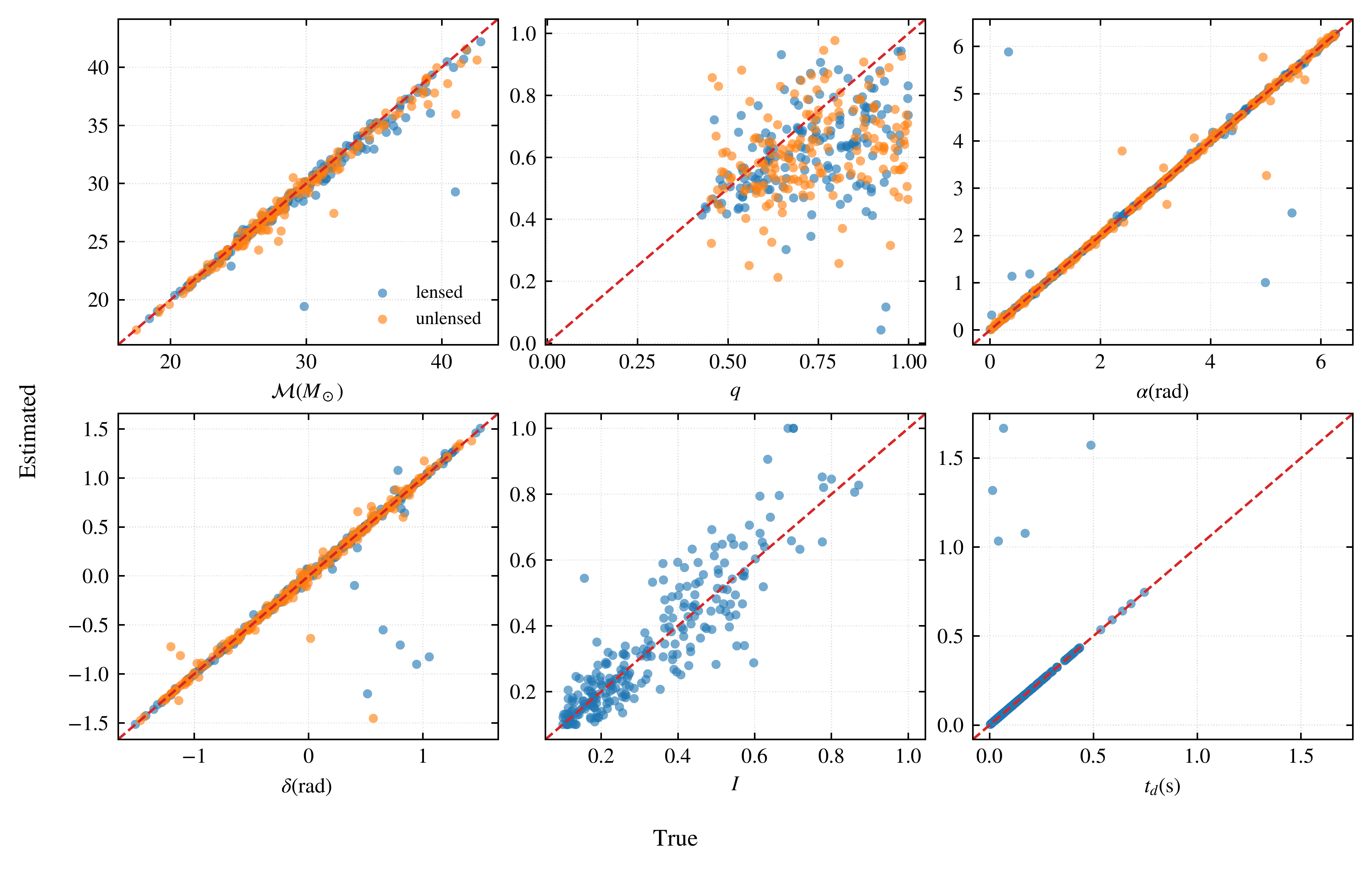}
    \caption{Parameter point estimation results of the best PSO run under the restricted 2-PN signal. The horizontal axis represents the true injected parameters, the vertical axis represents the PSO estimated parameters, and the dashed line indicates the ideal recovery case. The blue points and yellow points represent the best PSO estimation results corresponding to the lensed GW data and unlensed GW data, respectively.}
    \label{fig:pso_recovery_2pn}
\end{figure}

We first examine the parameter point estimation capability of PSO. Fig.~\ref{fig:pso_recovery_2pn} shows the relationship between the best PSO-run estimates and the true injected parameters for restricted 2-PN signals. In each subplot, the horizontal axis shows the true value, the vertical axis shows the PSO estimate, and the dashed line indicates ideal result. In the PSO search, $\tau_0$ and $\tau_{1.5}$ are used as the direct search parameters. In the subsequent analysis, these two
parameters are converted into compact-binary mass parameters using Eqs.~\eqref{tau_0},~\eqref{tau15} and ~\eqref{M}. The results in Fig.~\ref{fig:pso_recovery_2pn} show that, for the chirp mass $\mathcal{M}$, mass ratio $q=m_2/m_1$, sky-location parameters $(\alpha,\delta)$, and lensing parameters $(I,\Delta t_d)$, most samples lie close to the diagonal. This suggests that PSO can identify high-likelihood regions near the true injected parameters in the high-dimensional parameter space. Because unlensed GWs do not contain physically meaningful double-image lensing parameters, their recovery performance is evaluated primarily using $\mathcal{M}$, $q$, and the sky-location parameters.
% 我们首先考察PSO的参数估计能力。图~\ref{fig:pso_recovery_2pn}展示了在受限2-PN信号条件下，最佳PSO运行估计值与真实注入参数之间的关系。在每个子图中，横轴表示真实值，纵轴表示PSO估计值，虚线表示理想恢复情况。在PSO搜索中，$\tau_0$和$\tau_{1.5}$被用作直接搜索参数。在后续分析中，这两个参数通过式~\eqref{tau}和\eqref{M}转换为致密双的质量参数。图~\ref{fig:pso_recovery_2pn}的结果表明，对于啁啾质量$\mathcal{M}$、质量比$q=m_2/m_1$、天空定位参数$(\alpha,\delta)$和透镜参数$(I,\Delta t_d)$，大多数样本分布在对角线附近。这说明PSO能够在高维参数空间中识别出接近真实注入参数的高似然区域。由于未透镜化引力波不包含具有物理意义的双像透镜参数，因此其恢复性能主要通过$\mathcal{M}$、$q$和天空定位参数进行评估。

\begin{figure}
    \centering
    \includegraphics[width=\textwidth]{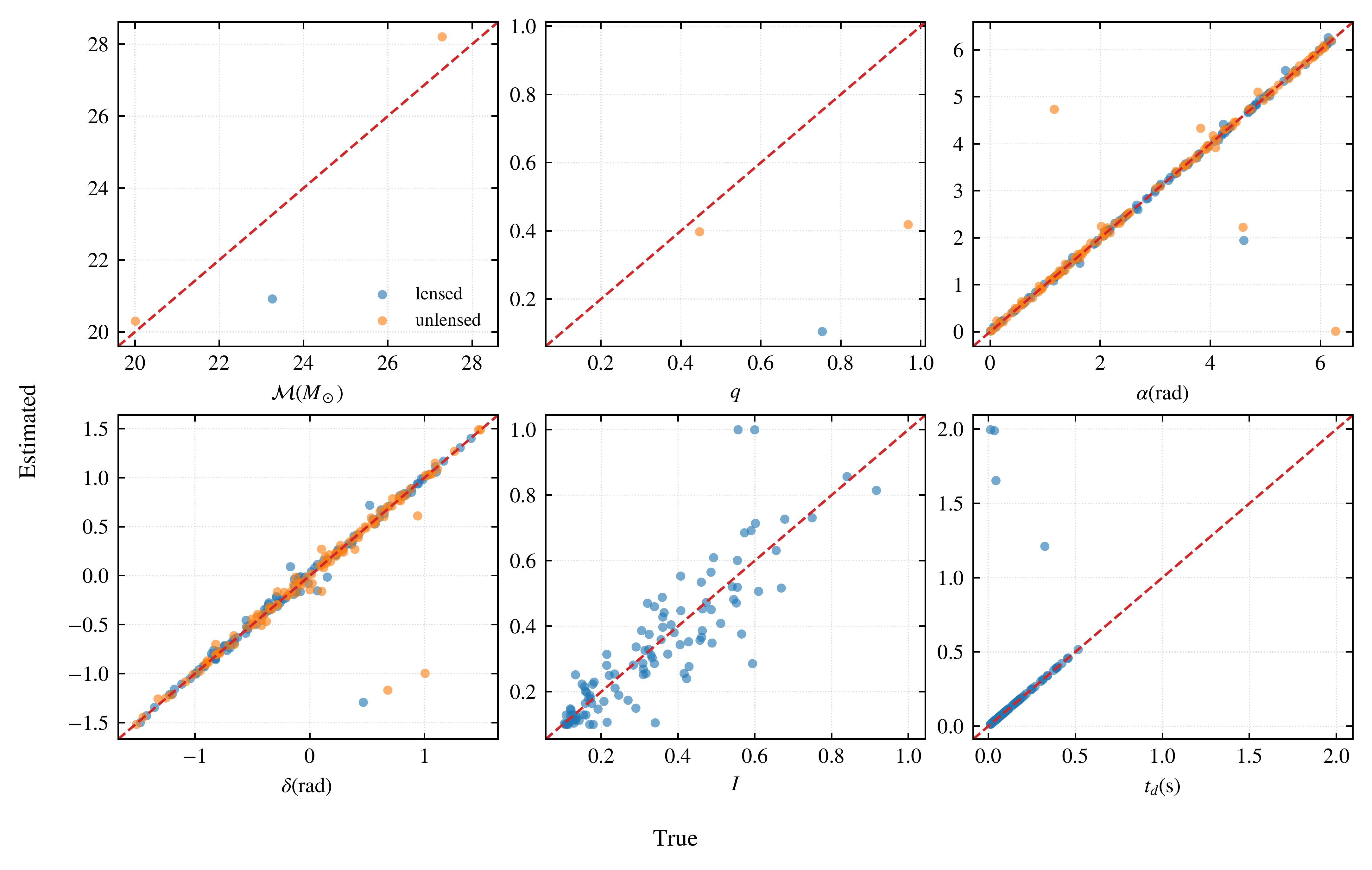}
    \caption{Parameter point estimation results of the best PSO run under the IMRPhenomD signal. These results are used to examine the robustness of the PSO search under the template–signal mismatch condition. The blue points and yellow points represent the best PSO estimation results corresponding to the lensed GW data and unlensed GW data, respectively.}
    \label{fig:pso_recovery_imrd}
\end{figure}

Figure~\ref{fig:pso_recovery_imrd} shows the parameter estimation results for IMRPhenomD signals when searched using the restricted 2-PN template. Compared with the restricted 2-PN results, the parameter estimates for IMRPhenomD exhibit relatively severe biases in the mass parameters, accurately reflecting the consequences arising from the signal–template mismatch. It should be noted that, in the mass parameter estimation of IMRPhenomD signals, since we invert $\mathcal{M}$ and $q$ from $\tau_0$ and $\tau_{1.5}$, a large number of samples yield unphysical mass-ratio inversion results; consequently, the effective points for $\mathcal{M}$ and $q$ in Fig.~\ref{fig:pso_recovery_imrd} are relatively few. In Fig.~\ref{fig:IMRD_tau}, we present the direct estimation results for the parameters $\tau_0$ and $\tau_{1.5}$. The results show that the majority of the PSO-estimated $\tau_0$ and $\tau_{1.5}$ values lie below the ideal line, which mainly originates from the mismatch between the signal and the template. Since IMRPhenomD contains more complete merger and ring-down phases, whereas the restricted 2-PN model we use is only an approximate description, such discrepancies in waveform information are to be expected. Nevertheless, the sky location parameters and lens-related parameters still exhibit good estimation results, indicating that PSO can still provide relatively reliable sky location information under non-ideal template conditions.
% 图~\ref{fig:pso_recovery_imrd}展示了IMRPhenomD信号在使用restricted 2-PN模板进行搜索时的参数估计结果。与 restricted 2-PN结果相比，IMRPhenomD的参数估计结果在质量参数方面存在较为严重偏差，准确反应出信号-模板失配带来的结果。需要指名的是，在IMRPhenomD信号的质量参数估计中，由于本文通过$\tau_0$和$\tau_{1.5}$反推出$\mathcal{M}$与$q$，导致大量样本出现非物理的质量比反推结果，因此图~\ref{fig:pso_recovery_imrd}中的$\mathcal{M}$和$q$有效点较少。在图~\ref{fig:IMRD_tau}中，我们给出了参数$\tau_0$和$\tau_{1.5}$的直接估计结果。结果显示，PSO估计得到的$\tau_0$和$\tau_{1.5}$值大多位于理想线之下，这主要源于信号与模板之间的失配。由于IMRPhenomD包含更完整的合并以及ring-down阶段，而本文所使用的restricted 2-PN模型仅为近似描述，因此产生这种波形信息上的误差是可以预期的。尽管如此，天空位置参数和透镜相关参数仍表现出良好的估计结果，说明PSO在非理想模板条件仍然能够提供相对可靠的天空位置信息。

\begin{figure}
    \centering
    \includegraphics[width=\textwidth]{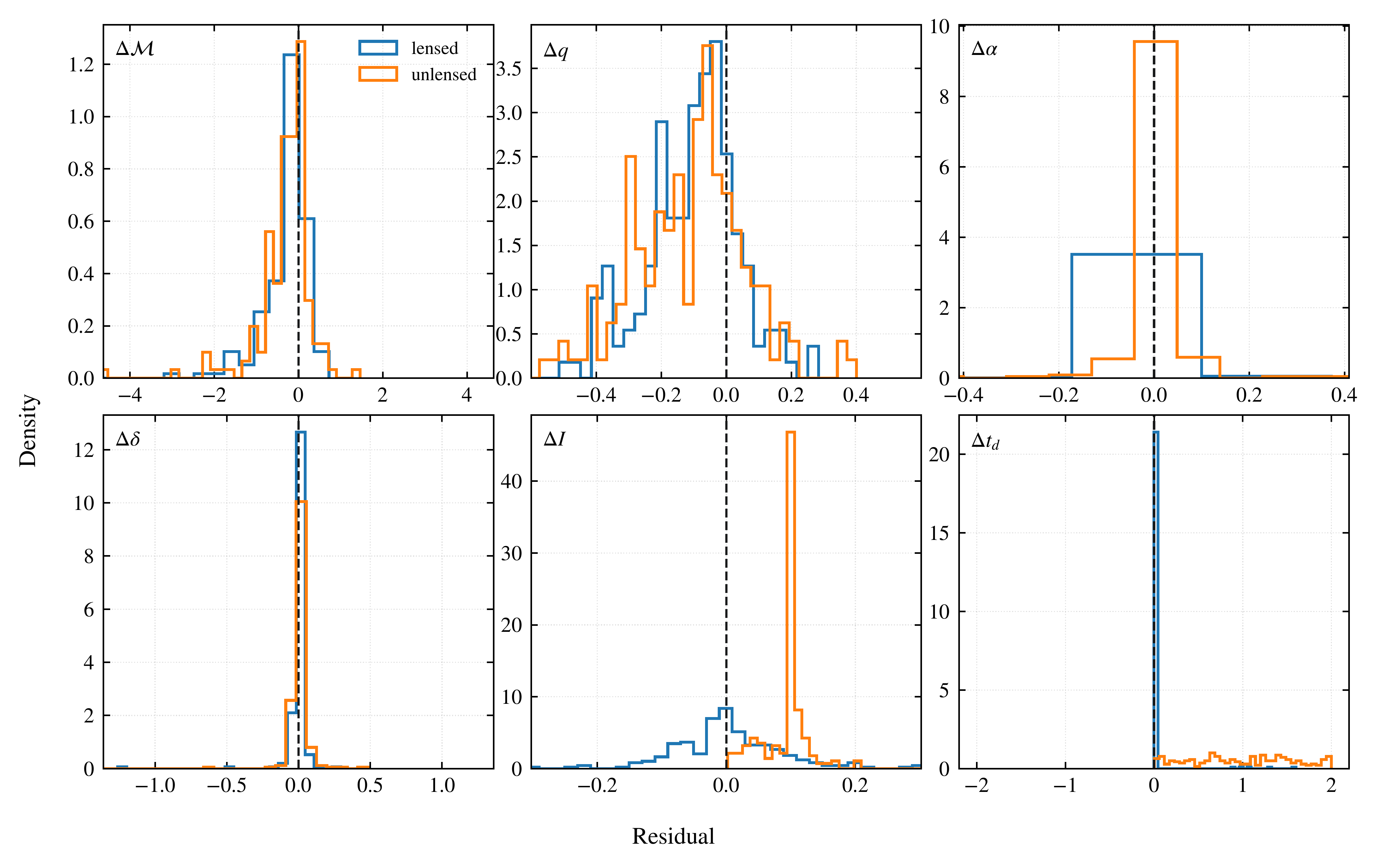}
    \caption{Residual distribution of PSO parameter estimation under the restricted 2-PN signal. The residual is defined as the difference between the estimated value and the true value.}
    \label{fig:pso_residual_2pn}
\end{figure}

To further illustrate the distribution of parameter estimation errors, Fig.~\ref{fig:pso_residual_2pn} presents the residual distribution between the estimated and true values under the restricted 2-PN signal condition. Overall, the residuals of most parameters are concentrated around zero, indicating that the PSO estimation results do not exhibit a significant global offset and that reliable parameter estimates can be obtained for the majority of the data. For the sky location parameters $(\alpha,\delta)$, the residual distribution is relatively concentrated, suggesting that the four-detector coherent network can provide effective constraints on sky localization. For lensed GW, the lens parameters $I$ and $t_d$ both exhibit good parameter estimation results. For unlensed data, since they do not inherently possess true lens parameters, we set both $I$ and $t_d$ to zero during data generation; therefore, for unlensed data, the errors in the lens parameters are essentially the estimated values themselves. It can be seen that the estimated values of the parameter $I$ are mostly concentrated around 0.1; we suspect that this serves as a discrimination threshold between lensed and unlensed cases, which may provide certain support for our subsequent signal classification. In contrast, $t_d$ exhibits results with a relatively high degree of dispersion. In general, PSO can provide reasonable parameter estimation results at a relatively low computational cost; however, even the best run results may still be affected by random initialization, multi-modal likelihood structures, and noise. Furthermore, the interference of noise may also cause pure noise data to be misclassified as a signal. Therefore, in the subsequent classification task, we do not directly rely on the best PSO output, but instead adopt a voting mechanism based on 12 independent runs to improve the stability of the results.
% 为了进一步展示参数估计误差的分布，图~\ref{fig:pso_residual_2pn}给出了restricted 2-PN信号条件下估计值与真实值之间的残差分布。整体来看，多数参数的残差均集中在零附近，说明PSO估计结果不存在显著的整体偏移，大部分的数据都能够给出可靠的参数估计结果。对于天空位置参数$(\alpha,\delta)$，残差分布较为集中，表明四探测器相干网络能够为天空定位提供有效约束。对于透镜化引力波信号，透镜参数$I$和$t_d$均表现出了良好的参数估计结果。而对于未透镜数据，由于其本身不具备真实的透镜参数，本文在数据生成时将$I$和$t_d$均设置为0，因此对于未透镜数据，透镜参数的误差实际就是估计值本身。可以看到，参数$I$的估计值大多集中在0.1左右，我们怀疑这是透镜与未透镜之间的一个分界阈值，可以为我们后续的信号分类提供一定的支撑。相较之下，$t_d$则展现出了离散程度较高的结果。总体而言，PSO能够在较低计算成本下给出合理的参数估计结果，但即使是最优的运行结果，仍可能受到随机初始化、多峰似然结构和噪声的影响。此外，噪声的干预还可能使纯噪声数据被误判为信号。因此，在后续分类任务中，本文不直接依赖最优PSO输出，而采用12次独立运行的投票机制提高结果稳定性。

\section{Classification}
\label{sect:classification}
In this classification task, we classify the PSO estimation results into three categories: lensed GW, unlensed GW, and noise-only samples. Because PSO is a stochastic optimization algorithm, a single run may be affected by the initial particle distribution, convergence to local optima, or the misidentification of noise as a signal. Therefore, classification based on a single PSO run can be unstable.

\begin{figure}[htbp]
    \centering

    \begin{subfigure}[b]{0.48\textwidth}
        \centering
        \includegraphics[width=\textwidth]{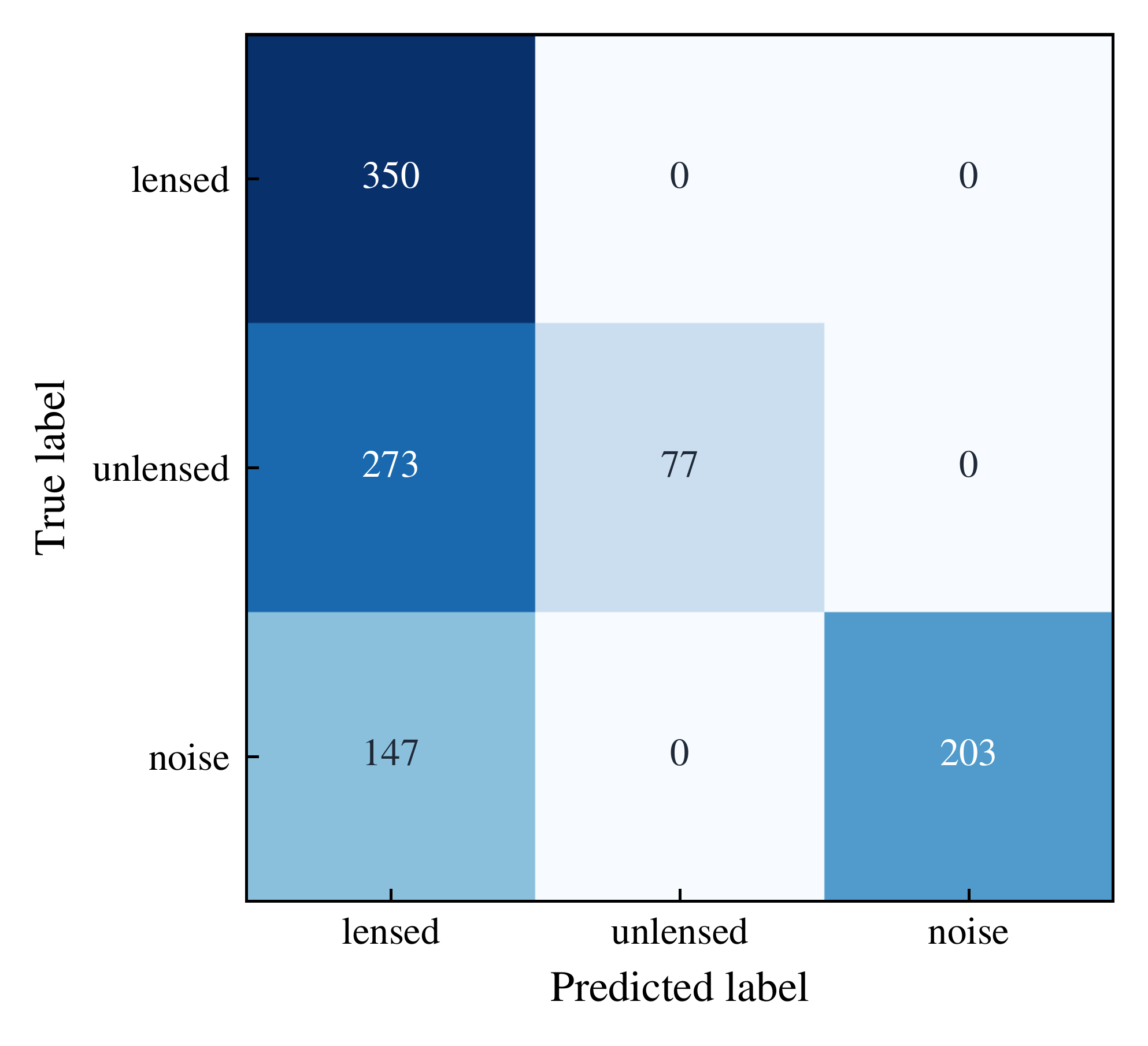}
        \caption{The baseline label confusion matrix obtained from signal classification using the best run result of 12 PSO runs, with a signal-to-noise ratio of 8 and $I = 0.05$ as the classification threshold.}
        \label{fig:pso_confusion_base}
    \end{subfigure}
    \hfill
    \begin{subfigure}[b]{0.48\textwidth}
        \centering
        \includegraphics[width=\textwidth]{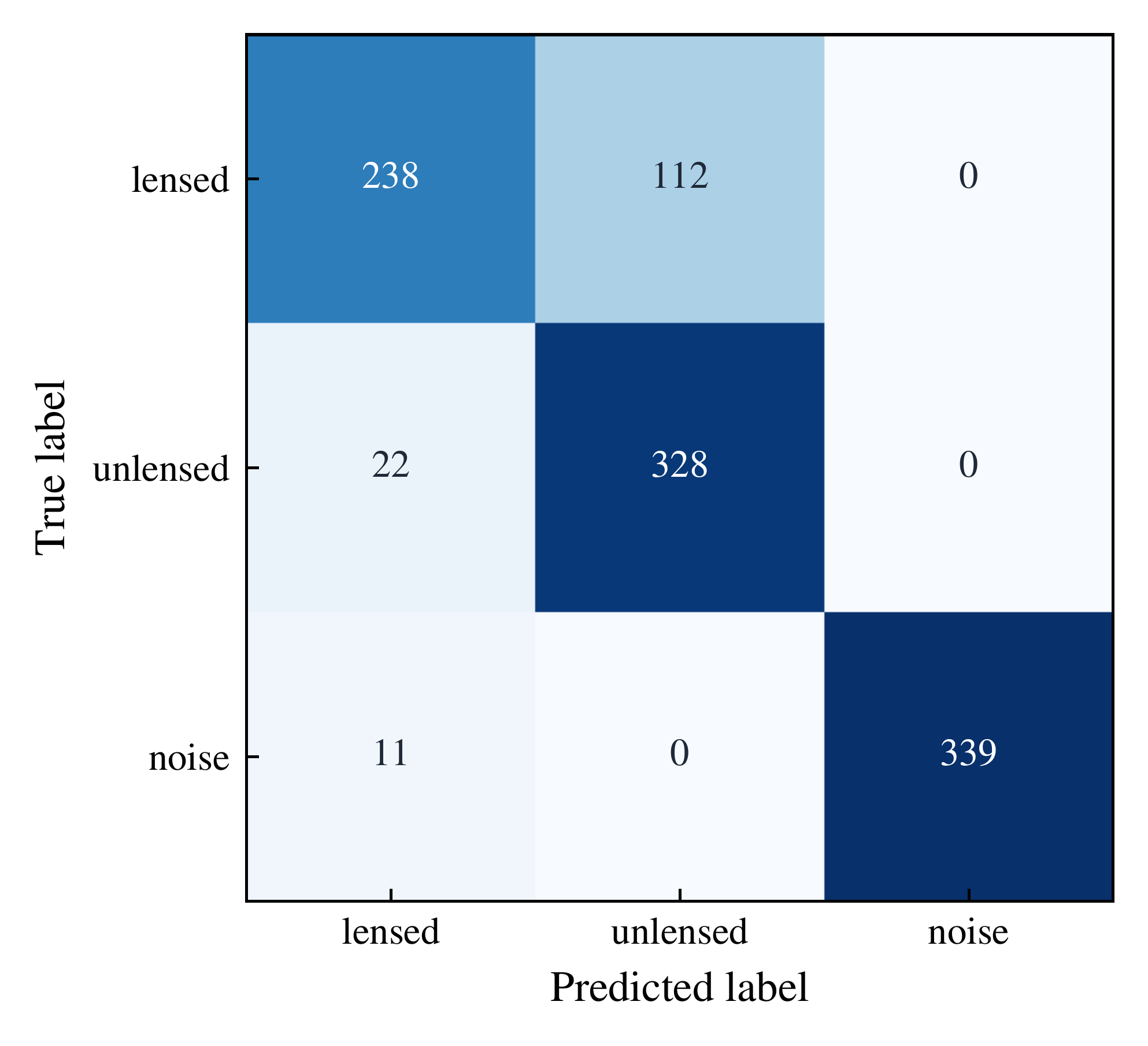}
        \caption{The three-class confusion matrix obtained from the voting mechanism and the $I_{\rm est}=0.11$ threshold discrimination based on the results of 12 independent PSO runs.}
        \label{fig:pso_confusion_vote}
    \end{subfigure}

    \caption{Confusion matrices obtained from the two different classification approaches}
    \label{fig:pso_flow_b_results}
\end{figure}

We first apply an initial threshold-based classification using a signal-to-noise ratio (SNR) threshold of 8 and an $I$ threshold of 0.05. However, this strategy yields low accuracy for the initial classification labels, as shown in Fig.~\ref{fig:pso_confusion_base}. Further analysis shows that the inferred $I$ values for most unlensed GW samples cluster around 0.1, indicating that $I=0.05$ is unsuitable for separating lensed and unlensed cases. In addition, for some noise-only samples, PSO misidentifies noise fluctuations as signals, producing SNR values slightly above 8 and thereby causing false signal classifications. These results indicate that both run-to-run variability and inappropriate thresholding can degrade classification performance. To mitigate these effects and incorporate information from all 12 independent runs, we construct a voting-based classification pipeline based on the multi-run results.
% 在分类任务中，我们实现了对PSO估计结果的信号三分类，即区分透镜化引力波信号、未透镜化引力波信号和噪声三类数据。由于PSO是随机优化算法，单次运行结果可能受到如初始粒子分布、陷入局部最优以及噪声误判为信号的影响，因此仅依赖一次PSO结果进行分类并不稳定。在初始分类中，我们发现使用信噪比为8，$I= 0.05$作为分类阈值得到的基础分类标签准确率非常低，结果如图~\ref{fig:pso_confusion_base}所示。通过分析发现，大多数的未透镜信号$I$集中在0.1左右，导致$I = 0.05$作为透镜与未透镜的分类阈值不够恰当。此外，有一定数量的纯噪声数据在PSO的搜索过程中，将噪声误认为信号，导致该次搜索的信噪比略高于8，从而将噪声误判为信号。为减弱单次运行的随机性带来的影响，并充分利用12次独立运行的信息，我们将分类任务调整为基于多次运行结果构建投票分类流程。

% 如图~\ref{fig:voting_scheme}所示，最终分类通过以下投票流程获得。首先，对12次独立PSO运行得到的基线分类标签进行多数投票，以判断样本是否为纯噪声。如果多数投票结果指示为噪声，则该样本被直接归为纯噪声类。其次，对于未被分类为噪声的样本，利用PSO估计参数$I_{\rm est}$进行二次判别。该参数具有物理动机，因为未透镜化引力波预期不会表现出显著的双像透镜调制，而透镜化引力波通常对应非零的像流量比。第三，基于参数估计误差分析，我们将阈值设为$I_{\rm est}=0.11$。对于每一次非噪声PSO运行，当$I_{\rm est}\geq0.11$时将其标记为透镜化引力波，当$I_{\rm est}<0.11$时将其标记为未透镜化引力波。最后，对更新后的非噪声标签再次进行多数投票，并将得到的多数标签作为最终信号类别。

Figure~\ref{fig:pso_confusion_vote} presents the three-class confusion matrix under the above ``voting + threshold" pipeline. The results show that the final three-class classification accuracy is $86.19\%$, 905 out of 1050 samples are correctly classified. For noise samples, 339 out of 350 are correctly identified, indicating that the voting mechanism can effectively distinguish noise from real GW and also confirming that PSO does indeed produce misclassifications. For unlensed GW, 328 out of 350 samples are correctly identified, indicating that the modified threshold can significantly suppress the misclassification of unlensed GW as lensed GW. For lensed GW, 238 out of 350 samples are correctly identified, with the main error arising from some lensed GW being classified as unlensed GW. This result suggests that the current threshold selection is more inclined to reduce the false-alarm rate of unlensed GW, but to some extent sacrifices the recall rate of lensed GW. Overall, these results indicate that the results of multiple independent PSO runs can not only be used for parameter estimation, but can also, when combined with a voting mechanism and physical thresholds, construct a relatively stable three-class classification pipeline and achieve satisfactory classification accuracy.
% 图~\ref{fig:pso_confusion_vote}给出了上述``投票+阈值"流程下的三分类混淆矩阵。结果显示，最终三分类准确率为$86.19\%$，即1050个样本中有905个被正确分类。对于噪声样本，350个样本中有339个被正确识别，说明投票机制能够有效区分噪声与真实引力波信号，也证明PSO确实会出现误判的情况。对于未透镜化信号，350个样本中有328个被正确识别，表明修改后的阈值可以显著抑制未透镜化信号被误判为透镜化信号的情况。对于透镜化信号，350个样本中有238个被正确识别，主要误差来自部分透镜化信号被判定为未透镜化信号。这说明当前阈值选择更倾向于降低未透镜化信号的误报，但会在一定程度上牺牲透镜化信号的召回率。总体而言，这一结果表明，PSO的多次独立运行结果不仅能够用于参数估计，还能够结合投票机制和物理阈值构建较为稳定的三分类流程，并取得较好的分类准确率。

Since a large fraction of the lensed signals in Fig.~\ref{fig:pso_confusion_vote} are classified as unlensed, we examine this subset in more detail. For weakly lensed signals with $I_{\mathrm{true}}<0.15$, the classification accuracy after voting is only 17.9\%, and every misclassified sample is labeled as unlensed; the accuracy rises to 62.1\% for $I_{\mathrm{true}}\in[0.15,0.2)$, 77.6\% for $I_{\mathrm{true}}\geq 0.15$, and 81.4\% for $I_{\mathrm{true}}\geq 0.2$. This indicates that the pipeline can barely detect weakly lensed signals with $I\lesssim 0.15$; the effective detection limit is approximately $I\approx 0.15$. We further examine the effect of adopting $I=0.15$ as the classification threshold: the overall accuracy after voting decreases from 86.19\% to 83.90\%, with the lensed GW accuracy dropping by 12.29\% and the unlensed GW accuracy rising by 5.43\%. Since all the misclassified weakly lensed samples are assigned to the unlensed class rather than the noise class, no false lensed candidates are produced; moreover, raising the threshold would only reduce the overall accuracy. We therefore keep $I=0.11$ as the classification threshold, which yields a higher overall accuracy.
% 由于图~\ref{fig:pso_confusion_vote}中有大量的透镜化引力波被判定为未透镜引力波，我们对该部分数据进行了进一步分析。对于 $I_{true}<0.15$ 的弱透镜化信号，投票后的分类正确率仅为 17.9\%，且漏判样本全部被分类为未透镜化；当 $I_{true}\in[0.15,0.2)$ 时正确率升至 62.1\%，$I_{true}\geq 0.15$ 后整体达到 77.6\%，当 $I_{true}\geq 0.2$ 后为 81.4\%。这表明当前流水线对 $I\lesssim 0.15$ 的弱透镜信号基本不可检出，其有效检测下限约为 $I\approx 0.15$。我们同时检验了以 $I=0.15$ 作为分类阈值的效果：投票后总正确率由 86.19\% 降至 83.90\%，其中 lensed GW 分类正确率降低 12.29\%，unlensed GW 提高 5.43\%。由于漏判的弱透镜样本全部被归类为未透镜而非噪声，且提高阈值反而使总正确率下降，因此本文仍采用总正确率更高的 $I=0.11$ 作为分类阈值。

\begin{figure}[htbp]
  \centering
  \includegraphics[width=\textwidth]{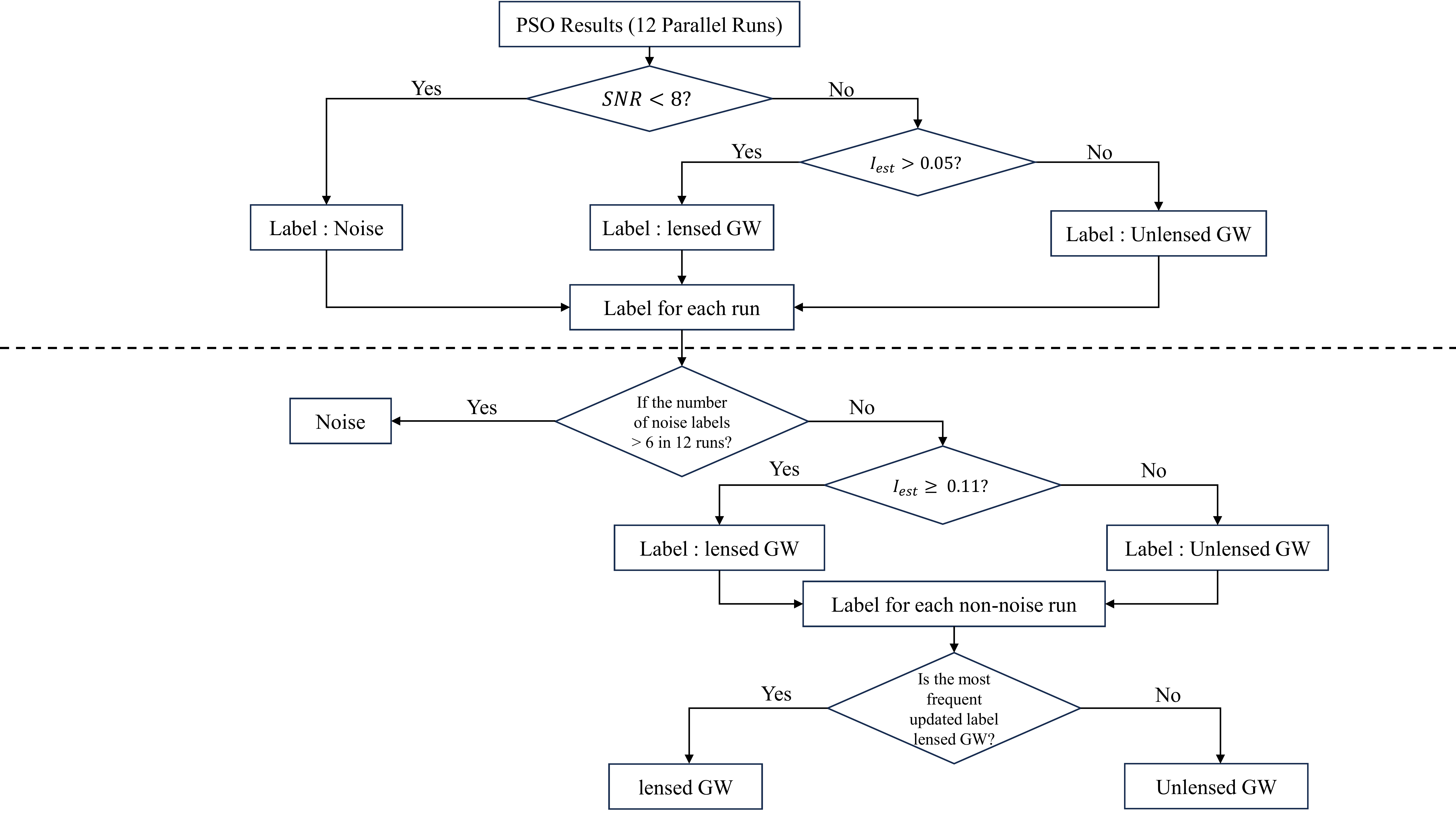}
  \caption{Flowchart of the two-stage majority-voting classification scheme based on 12 independent PSO runs. In the first voting stage, the most frequent label is used to  determines whether the sample is noise. For samples not classified as noise, the runs labeled as non-noise are relabeled using the threshold $I_{\rm est} = 0.11$, and the most frequent updated label is assigned as the final lensed or unlensed GW classification.}
  \label{fig:voting_scheme}
\end{figure}

As illustrated in Fig.~\ref{fig:voting_scheme}, the final classification is determined based on the labels obtained from 12 prarllel PSO runs. In this procedure, we implement hard voting by selecting the most frequent label among the runs:
\begin{enumerate}
\item For each of the 12 independent PSO runs, an initial label is assigned based on the estimated SNR and the PSO-estimated parameter $I_{\rm est}$. If ${\rm SNR}<8$, the run is labeled as noise. Otherwise, the run is further classified according to $I_{\rm est}$: it is labeled as lensed GW if $I_{\rm est}>0.05$, and unlensed GW if $I_{\rm est}\leq0.05$. 
\item After all runs are labeled, the first voting step selects the most frequent label. if the most frequent label is noise, the sample is assigned to the noise-only class.
\item For samples not classified as noise, only the runs intially labeled as non-noise are retained for further discrimination. These retained runs are relabeled using a threshold of $I_{\rm est}=0.11$, motivated by the parameter-estimation error analysis. Specifically, each remaining runs is relabeled as lensed GW if $I_{\rm est} \geq 0.11$ and as an unlensed GW when $I_{\rm est}< 0.11$.
\item Finally, a second voting step is applied to the updated lensed/unlensed labels. The most frequent updated label is assigned as final signal category.
\end{enumerate}

\begin{figure}[htbp]
  \centering
  \includegraphics[width=\textwidth]{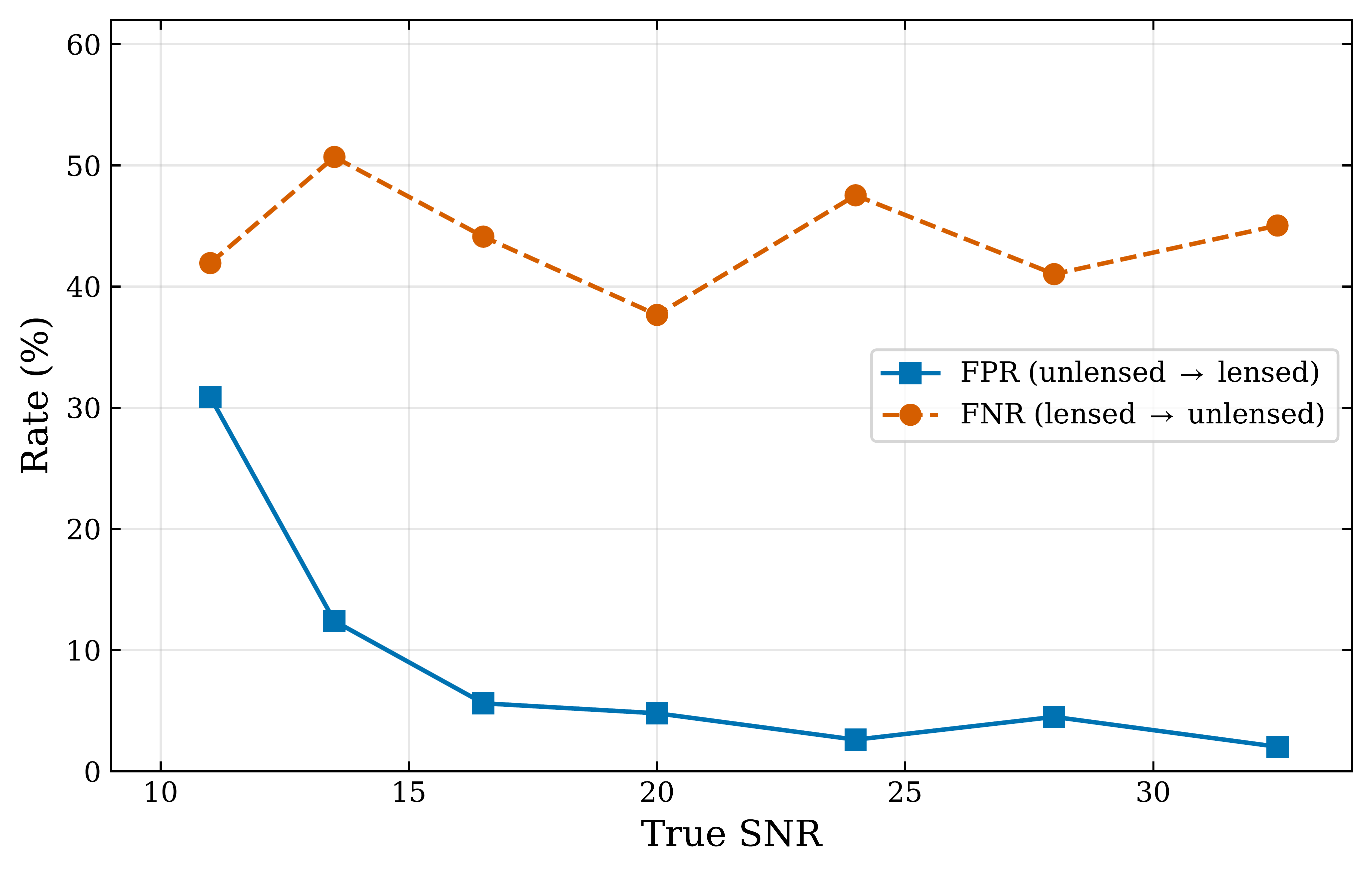}
  \caption{The impact of SNR on lens signal classification，The brown curve in the figure shows the probability of mistaking an unlensed GW for a lensed one, while the blue curve shows the probability of correctly identifying a lensed GW.}
  \label{fig:SNR_FPR}
\end{figure}

We use a fixed threshold $I_{\mathrm{est}}=0.11$ to distinguish lensed from unlensed gravitational-wave signals. To examine the influence of noise on the classification, we group the samples into SNR bins and compute the misclassification rates in each bin, as shown in Fig.~\ref{fig:SNR_FPR}. The FPR(False Positive Rate) is as high as 30.9\% for SNR $\in$ [10, 12) and decreases overall with increasing SNR, dropping to about 5\% for SNR $>$ 15 and to only about 2\%--4\% for SNR $>$ 25. Meanwhile, the FNR(False Negative Rate)) stays within 38\%--51\% across all SNR bins, showing no significant dependence on the SNR. Furthermore, we perform a threshold scan by maximizing Youden's J index and find that the optimal threshold can be well approximated by $I_{th}(\mathrm{SNR}) \approx 0.08 + 0.50/\mathrm{SNR}$. Nevertheless, we retain the fixed threshold of 0.11 for two reasons. First, the SNR-independence of the FNR implies that an SNR-dependent threshold cannot improve the recovery of weakly lensed signals. Second, the recovered SNR is itself subject to estimation uncertainty in real observations, so conditioning the threshold on the SNR would introduce additional uncertainty. We therefore adopt the fixed threshold $I=0.11$, and rely on the voting mechanism to improve the overall classification accuracy.

\section{Posterior distribution }%Posterior
\label{sect:psoGaussMix}
In addition to the search and classification capabilities of PSO, we further investigate the performance of PSO combined with nested sampling. Nested sampling is a Bayesian sampling method that estimates parameter posterior distributions and credible regions. However, in high-dimensional parameter spaces, directly adopting
broad priors often makes nested sampling computationally expensive because the posterior region must be explored adequately. In contrast, PSO can rapidly locate
high-likelihood regions at lower computational cost, but it primarily provides optimization-based point estimates and cannot directly produce a full Bayesian
posterior distribution.

Based on this complementarity, we use the rapid search capability of PSO and adopt neighborhoods around its point estimates as priors for nested sampling.
Specifically, we use the candidate parameter points obtained from multiple PSO runs to construct a Gaussian-mixture prior within a $3\sigma$ range, where $\sigma$ is estimated from the standard deviations of these point estimates. We denote this prior as PSO\_GAUSS\_MIX. We then compare the results with nested-sampling results obtained using the uniform priors listed in Table~\ref{tab:parameter search range}. The core idea is that PSO first rapidly locates high-likelihood candidate regions in the global parameter space. Nested sampling is then performed under the PSO\_GAUSS\_MIX prior constraints, improving sampling efficiency while producing parameter posterior distributions.To ensure that the prior does not extend into unphysical regions of the parameter space, truncation and boundary ranges are imposed during the construction of the PSO\_GAUSS\_MIX prior. Specifically, each Gaussian component centered at a candidate point is truncated at $3\sigma$: the probability density is set to zero wherever $|x-\mu|>3\sigma$ and retained elsewhere, where $\sigma$ denotes the standard deviation of the candidate-point estimates. For the periodic parameter $\alpha$, the distance is measured in a wrapped sense, so that the $3\sigma$ neighborhood crossing the boundary wraps around to the other side without being truncated. The truncated Gaussian mixture is then implemented as an interpolated prior whose domain is explicitly restricted to the physical ranges of the parameters, with the probability density strictly zero outside these ranges; $\alpha$ is additionally assigned a periodic boundary to ensure continuity across $0/2\pi$. The resulting prior is the PSO\_GAUSS\_MIX prior.

\begin{figure}
    \centering
    \includegraphics[width=\textwidth]{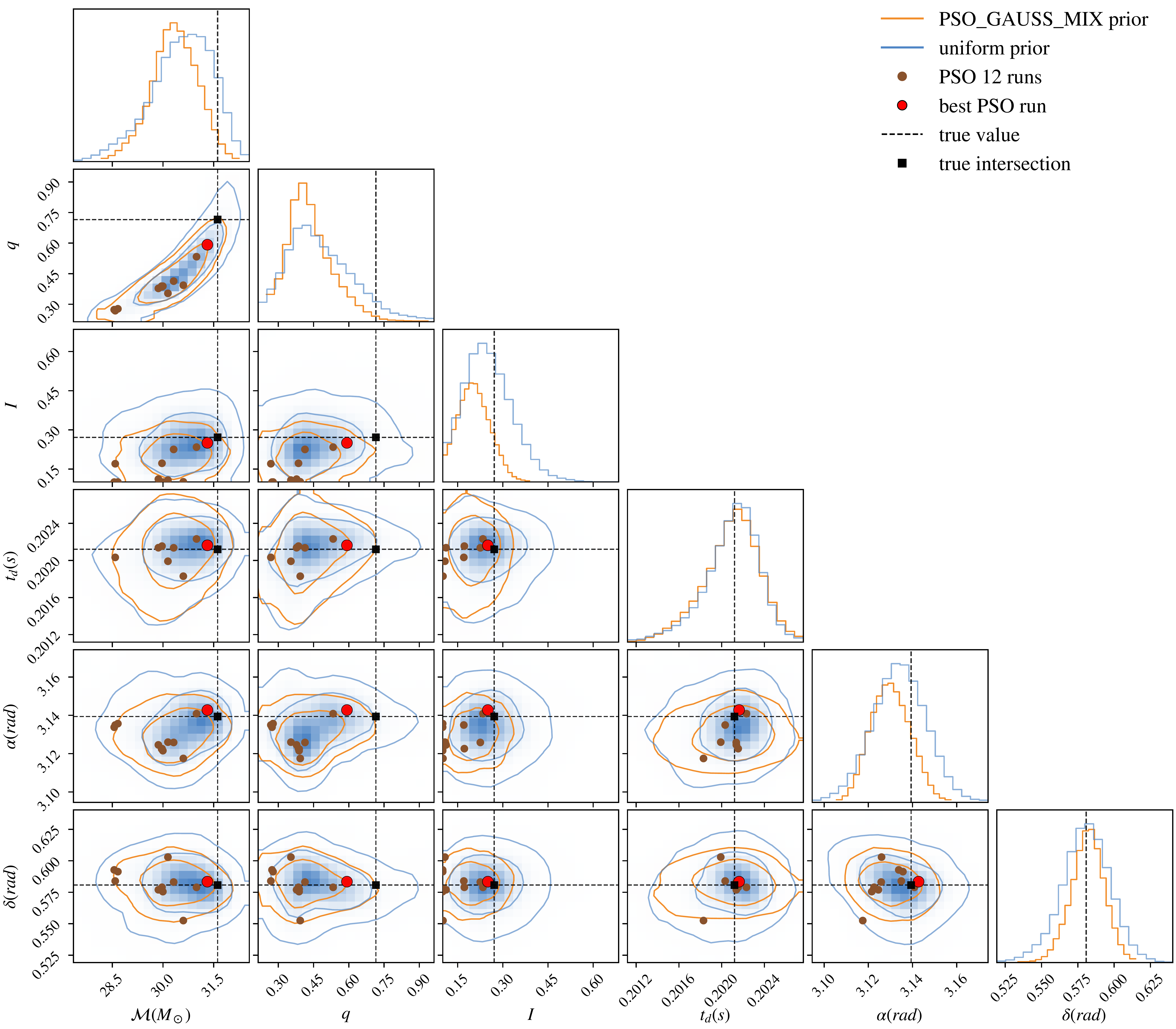}
    \caption{nested sampling posterior distribution plots for uniform and PSO\_GAUSS\_MIX. The contours represent the 95\% and 68\% credible regions, with the blue contours corresponding to the posterior distribution of uniform and the yellow contours corresponding to the posterior distribution of PSO\_GAUSS\_MIX. The brown points denote the results of multiple PSO runs, the red points denote the best run results of the 12 PSO runs, and the black markers indicate the true injected parameters.}
    \label{fig:pso_gauss_mix_corner}
\end{figure}

Figure~\ref{fig:pso_gauss_mix_corner} shows the comparison of the nested sampling posterior distributions under uniform and PSO\_GAUSS\_MIX. In the figure, the blue contours represent the posterior credible regions under uniform, and the yellow contours represent the posterior credible regions under PSO\_GAUSS\_MIX. The contours represent the 95\% and 68\% credible intervals, respectively. The black dashed lines and markers indicate the true injected parameter positions, the green points represent the results of multiple PSO runs, and the red points represent the best PSO run results. It can be seen that the PSO run results are typically distributed near the high-probability posterior regions of uniform, indicating the accuracy of the PSO results. Compared with uniform, the posterior regions under PSO\_GAUSS\_MIX are generally more concentrated, yielding more accurate posterior distributions. This indicates that PSO-informed does not constrain nested sampling to incorrect regions, but rather effectively shrinks the search space while preserving the main high-credibility regions.
% 图~\ref{fig:pso_gauss_mix_corner}展示了uniform与PSO\_GAUSS\_MIX下nested sampling后验分布的对比。图中蓝色等高线表示uniform  的后验置信区域，黄色等高线表示PSO\_GAUSS\_MIX的后验置信区间。等高线分别代表95\%和68\%置信区间。黑色虚线和标记表示真实注入参数位置，绿色点表示多次PSO运行结果，红色点表示PSO最优运行结果。可以看出，PSO运行结果通常分布在uniform  的后验高概率区域附近，表明PSO结果的准确性。与uniform  相比，PSO\_GAUSS\_MIX  下的后验区域通常更加集中，能够得到更加准确的后验分布。这表明PSO-informed  并未将nested sampling限制在错误区域，而是在保留主要高置信区域的同时有效收缩了搜索空间。

These results demonstrate that PSO\_GAUSS\_MIX offers two advantages. First, compared with using PSO alone, it can provide the complete posterior distribution and credible regions through nested sampling, thereby compensating for PSO's lack of posterior uncertainty quantification. Second, compared with nested sampling directly using uniform, it leverages the high-likelihood candidate regions provided by PSO to compress the prior space, thereby reducing ineffective sampling in low posterior probability regions. Therefore, PSO\_GAUSS\_MIX can be regarded as an effective fusion strategy that combines rapid global optimization with Bayesian posterior estimation.

\begin{table}[htbp]
\centering
\caption{Parameter summary for FIG. ~\ref{fig:pso_gauss_mix_corner}}
\label{tab:sample0014_posterior_summary}
\begin{tabular}{lcccccc}
\hline
 & $\mathcal{M}(M_\odot)$ & $q$ & $I$ & $t_d$(s) & $\alpha$(rad) & $\delta$(rad) \\
\hline
True value          & 31.62 & 0.7153 & 0.2714 & 0.2021 & 3.139 & 0.5807 \\
Posterior(uniform prior)     & $30.64_{-0.96}^{+0.78}$ & $0.4552_{-0.10}^{+0.17}$ & $0.2435_{-0.06}^{+0.08}$ & $0.2021_{-0.00}^{+0.00}$ & $3.134_{-0.01}^{+0.01}$ & $0.5812_{-0.02}^{+0.01}$ \\
Posterior(PSO\_GAUSS\_MIX)  & $30.35_{-0.63}^{+0.67}$ & $0.4193_{-0.06}^{+0.11}$ & $0.2037_{-0.05}^{+0.05}$ & $0.2021_{-0.00}^{+0.00}$ & $3.13_{-0.01}^{+0.01}$ & $0.582_{-0.01}^{+0.01}$ \\
PSO + $1\sigma$ Gaussion  &$30.01_{-1.28}^{+1.25}$ & $0.3442_{-0.01}^{+0.17}$ &$0.1437_{-0.07}^{+0.09}$ &$0.4214_{-0.28}^{+0.46}$ &$3.129_{-0.01}^{+0.01}$ &$0.582_{-0.01}^{+0.01}$\\
\hline
\end{tabular}
\tablecomments{1\textwidth}{
	The rows for uniform and PSO\_GAUSS\_MIX present the corresponding posterior distributions, while the PSO + 1$\sigma$  Gaussian distribution is a superposition of Gaussians with 1$\sigma$ intervals from 12 PSO results.}
\label{tab:Posterior parameter summary}
\end{table}

Table~\ref{tab:Posterior parameter summary} further presents the quantitative posterior parameter results for the sample shown in Fig.~\ref{fig:pso_gauss_mix_corner}, where each parameter is given in the form of the posterior median and the 68\% credible interval. It can be seen that the parameter estimates obtained under PSO\_GAUSS\_MIX and uniform are broadly consistent. For the time delay $t_d$ and the sky location parameters $\alpha$ and $\delta$, the posterior medians under the two priors are nearly identical, and the credible interval widths are also relatively close. For the parameters $\mathcal{M}$, $q$, and $I$, both the PSO\_GAUSS\_MIX and uniform results exhibit certain deviations from the true values, but they are largely consistent with each other. Meanwhile, the credible intervals under PSO\_GAUSS\_MIX are narrower than those under uniform, indicating that the prior information provided by PSO helps nested sampling further shrink the posterior distribution and improve the concentration of parameter estimation.

Fig.~\ref{fig:appendix_pso_prior_posterior_comparison} further compares the PSO+$1\sigma$ Gaussian parameter distribution with the posterior distributions obtained from nested sampling under the uniform and PSO\_GAUSS\_MIX priors. Unlike Fig.~\ref{fig:pso_gauss_mix_corner}, this supplementary plot explicitly shows the Gaussian distribution reconstructed from the 12 PSO point estimates, with the Gaussian width set by their $1\sigma$ scatter. The PSO estimates are broadly located near the posterior high-probability region, confirming the effectiveness of PSO in the search phase. However, since PSO provides discrete optimization outputs, the reconstructed PSO+$1\sigma$ Gaussian distribution cannot fully reproduce the posterior morphology. For parameters such as $\mathcal{M}$, $q$, and $I$, its center or width can deviate from the posterior because a few PSO runs may differ from the others and broaden or shift the reconstructed distribution. Therefore, this distribution should be regarded as an auxiliary representation of the candidate high-likelihood regions identified by PSO, rather than as a substitute for the posterior distribution. Nevertheless, the posterior distributions obtained under the uniform and PSO\_GAUSS\_MIX priors remain broadly consistent in the main high-probability region, while PSO\_GAUSS\_MIX can yield more concentrated posterior constraints. This indicates that PSO provides useful prior information for nested sampling, whereas nested sampling produces the full Bayesian posterior distribution.

Finally, we compare the average runtime of nested sampling based on uniform and PSO\_GAUSS\_MIX prior, with the results shown in Table~\ref{tab:runtime}. $t_1$ is the average PSO runtime for the 1050 data, and $t_2$ and $t_3$ represent the average PSO runtime for the restricted 2-PN model and the IMRPhenomD model under the lensed GW data condition, respectively. $t_4$–$t_7$ represent the average nested sampling runtime for the lensed GW data of the restricted 2-PN and IMRPhenomD models under the two different prior distribution, uniform and PSO\_GAUSS\_MIX, respectively. From the results of $t_4$–$t_7$, for the restricted 2-PN model, the average sampling time under uniform is $400.531\,{\rm s}$, whereas that under PSO\_GAUSS\_MIX is $140.788\,{\rm s}$, corresponding to a speedup of approximately $2.84\times$. For the IMRPhenomD model, the average sampling time under uniform is $427.310$\,s, and that under PSO\_GAUSS\_MIX is $206.690\,{\rm s}$, corresponding to a speedup of approximately $2.07\times$. $t_2 + t_6$ represents the total time cost of using PSO and PSO\_GAUSS\_MIX combined, with a total duration of $289.545$, which is still lower than the runtime of uniform, yielding a speedup of approximately $1.38\times$.

\begin{table}[htbp]
\centering
\caption{Comparison of mean running time under different methods and model}
\label{tab:mean_time_comparison}
\begin{tabular}{ccccc}
\toprule
    Method & Prior  & Model & Mean time (s) & Mark \\
\midrule
\multirow{3}{*}{PSO}
    & - & -  & 152.47 &$t_1$ \\
    & - & restricted 2-PN  & 148.757&$t_2$ \\
    & - & IMRPhenomD & 147.019 &$t_3$\\
\midrule
\multirow{4}{*}{nested sampling}
    & uniform        & restricted 2-PN  & 400.531 &$t_4$\\
    & uniform        & IMRPhenomD & 427.31  &$t_5$\\
    & PSO\_GAUSS\_MIX & restricted 2-PN  & 140.788 &$t_6$\\
    & PSO\_GAUSS\_MIX & IMRPhenomD & 206.69  &$t_7$\\
\bottomrule
\end{tabular}
\label{tab:runtime}
\end{table}

The above results indicate that PSO\_GAUSS\_MIX can significantly reduce the computational cost of nested sampling. In particular, in the restricted 2-PN case, the candidate regions provided by PSO exhibit good consistency with the true posterior regions, and thus the improvement in sampling efficiency is most pronounced. In the IMRPhenomD case, although the constraining effect of PSO-informed is somewhat weakened due to the template–signal mismatch, it still yields an average speedup of more than twofold. Combined with the posterior distribution comparison results, it can be seen that PSO\_GAUSS\_MIX can still maintain coverage of the high-confidence posterior regions while improving computational efficiency. Therefore, in high-dimensional lensed GW parameter estimation tasks, PSO\_GAUSS\_MIX may offer an effective approach that balances search efficiency with posterior credibility.

\section{Conclusion and Discussion}
\label{sect:conclusion}
We investigate the use of PSO for the detection, parameter estimation, and classification of GW signals from lensed CBC. To address the high dimensionality of the parameter space, the multimodal nature of the likelihood surface, and the computational cost of conventional matched-filtering and Bayesian-sampling methods for lensed GW searches, we develop a PSO-based search pipeline within a fully coherent four-detector network. We characterize the lensing modulation using the model-independent double-image lensing parameters $(I, t_d)$, thereby reducing dependence on specific lens models. Using simulated data, we first evaluate the ability of PSO to recover the parameters of lensed GW signals modeled with a restricted 2PN waveform. The results show that PSO can rapidly identify high-likelihood regions over a broad parameter space and yield relatively stable parameter estimates. For unlensed GW signals, the PSO-estimated flux ratio $I$ tends to cluster at low values, providing a useful criterion for distinguishing lensed from unlensed signals. However, when a restricted 2PN template is used to search for IMRPhenomD signals, the recovered mass parameters exhibit substantial bias because of template–signal mismatch. This finding indicates that waveform-model inaccuracies can degrade the parameter-estimation performance of PSO and that the restricted 2PN model is insufficient to describe the full GW waveform. In addition to parameter estimation, we develop a three-class classification pipeline based on multiple independent PSO runs to distinguish among lensed GW signals, unlensed GW signals, and noise-only samples. To mitigate the effects of random initialization and local optima in individual PSO runs, we employ a majority-voting mechanism. We then apply a secondary decision rule based on the flux-ratio threshold $I_{\rm est}=0.11$. Together, these steps yield an overall classification accuracy of $86.19\%$ on 1,050 samples. Classification of noise-only samples and unlensed GW signals is relatively reliable, whereas most errors for lensed GW signals arise when weakly lensed samples are misclassified as unlensed. These results suggest that the current classification strategy prioritizes avoiding the false classification of unlensed GW signals as lensed, but at the cost of reduced recall for some lensed signals. Further analysis of weakly lensed signals shows that the current pipeline struggles to identify signals with $I \leq 0.15$, and the effective detection limit is around $I\approx0.15$. All missed samples are classified as unlensed signals rather than noise.

Furthermore, we investigate a hybrid PSO–nested sampling framework for rapid and accurate posterior inference. PSO can quickly identify high-likelihood candidate regions, but it does not provide a full characterization of posterior uncertainty. In contrast, nested sampling yields reliable Bayesian posterior distributions, but it is computationally expensive under broad priors. To exploit the complementary strengths of the two methods, we use 12 independent PSO point estimates as mixture centers. We then construct a Gaussian mixture prior covering a $3\sigma$ neighborhood around each center and incorporate it into nested sampling. The resulting method is denoted as PSO\_GAUSS\_MIX. The results show that this prior effectively compresses the parameter space, substantially improving sampling efficiency while preserving coverage of the main high-confidence posterior regions. Relative to a uniform prior, PSO\_GAUSS\_MIX yields speedups of approximately $2.84\times$ for the restricted 2-PN case and $2.07\times$ for the IMRPhenomD case. Even after accounting for the computational cost of the PSO stage, the total runtime remains lower than that of nested sampling performed directly under a uniform prior.

In summary, PSO demonstrates strong global search capability and low computational cost in searches for lensed GWs. It can serve both as a rapid tool for parameter estimation and candidate-event screening and, when combined with Bayesian sampling methods, as a means of constructing more informative prior distributions. Our results further suggest that incorporating stochastic global optimization methods into lensed GW searches can achieve a favorable balance between computational efficiency and parameter recovery performance. These methods may therefore provide a practical alternative for the rapid screening of lensed GW signals in future large-scale GW datasets.

Future work can proceed along three main directions. First, our main search template is based on the restricted 2-PN waveform model, which does not fully capture key physical effects, including merger and ring-down, spin, eccentricity, precession, and higher-order modes. Future studies should therefore extend the search templates to more complete waveform families, such as IMRPhenom (\citealt{MateuLucena+etal+2022,Pratten+etal+2021}) and SEOBNR (\citealt{Cotesta+etal+2020}). They should also systematically compare the stability of PSO in recovering mass parameters, sky localization, and lensing parameters across different waveform models. In particular, our IMRPhenomD injection experiments show that template–signal mismatch can bias the recovered mass parameters. More accurate waveform modeling is therefore essential for improving parameter-estimation precision. Second, we use the model-independent double-image parameters $(I,t_d)$ to describe the lensing effect. This parameterization is suitable for characterizing double-image interference in the geometric-optics regime, but it does not yet cover more complex scenarios, such as galaxy-scale lenses and wave-optics effects, nor does it incorporate more accurate computational treatments (\citealt{Guo+Lu+2020,Tambalo+etal+2023}). In addition, wave-optics effects may be non-negligible for low-mass lenses or low-frequency signals. Incorporating the frequency-dependent lens amplification factor into the PSO search framework is therefore an important topic for future investigation (\citealt{Grespan+Biesiada+2023}). Furthermore, the searchs in this work are perfromed under Guassion-noise conditions, and the potential impact of non-Gaussion noise not yet been examined. In more realistic data, non-Gaussian noise features may influence both the point estimates and the inferred posterior distributions. Future work we will quantify how non-Gaussion noise affects the accuracy, stability, and robustness of PSO and PSO\_GAUSS\_MIX. Third, at the algorithmic level, PSO can be integrated with methods beyond nested sampling. As GW observatories continue to advance—including LVK (\citealt{Abbott+etal+2018b,Abbott+etal+2021}), LISA (\citealt{Barausse+etal+2020}), Taiji (\citealt{Hu+Wu+2017,Ruan+etal+2020}), and TianQin (\citealt{Luo+etal+2016,Luo+etal+2025})—both the number and the complexity of future GW events are expected to increase, along with the likelihood of detecting lensed events. Accordingly, a key challenge will be to develop automated search pipelines that balance computational efficiency, parameter-estimation accuracy, and statistical rigor. The PSO-based search and PSO-informed Bayesian inference framework proposed in this work provides a preliminary step in this direction. It could also be integrated with deep-learning architectures by leveraging the low-cost exploratory capability of PSO for hyperparameter optimization (\citealt{Benedetto+etal+2023,Schafer+etal+2023,Goyal+etal+2021,Qin+etal+2025}). With more realistic data-processing pipelines, more complete physical models, and more efficient hybrid optimization and posterior-inference techniques, this framework could be applied to the rapid screening, parameter estimation, and subsequent astrophysical interpretation of large samples of lensed GW candidates.

\normalem
\begin{acknowledgements}
This work is supported by the National Natural Science Foundation of China under Grant No.~12503001 (X.G.) and Shandong Provincial Natural Science Foundation under Grant No.~ZR2024QA219 (M.Z.).

\end{acknowledgements}

\bibliographystyle{raa}
\bibliography{bibtex}

\newpage
\appendix
\section{}
\begin{figure}[htbp]
    \centering
    \includegraphics[width=\textwidth]{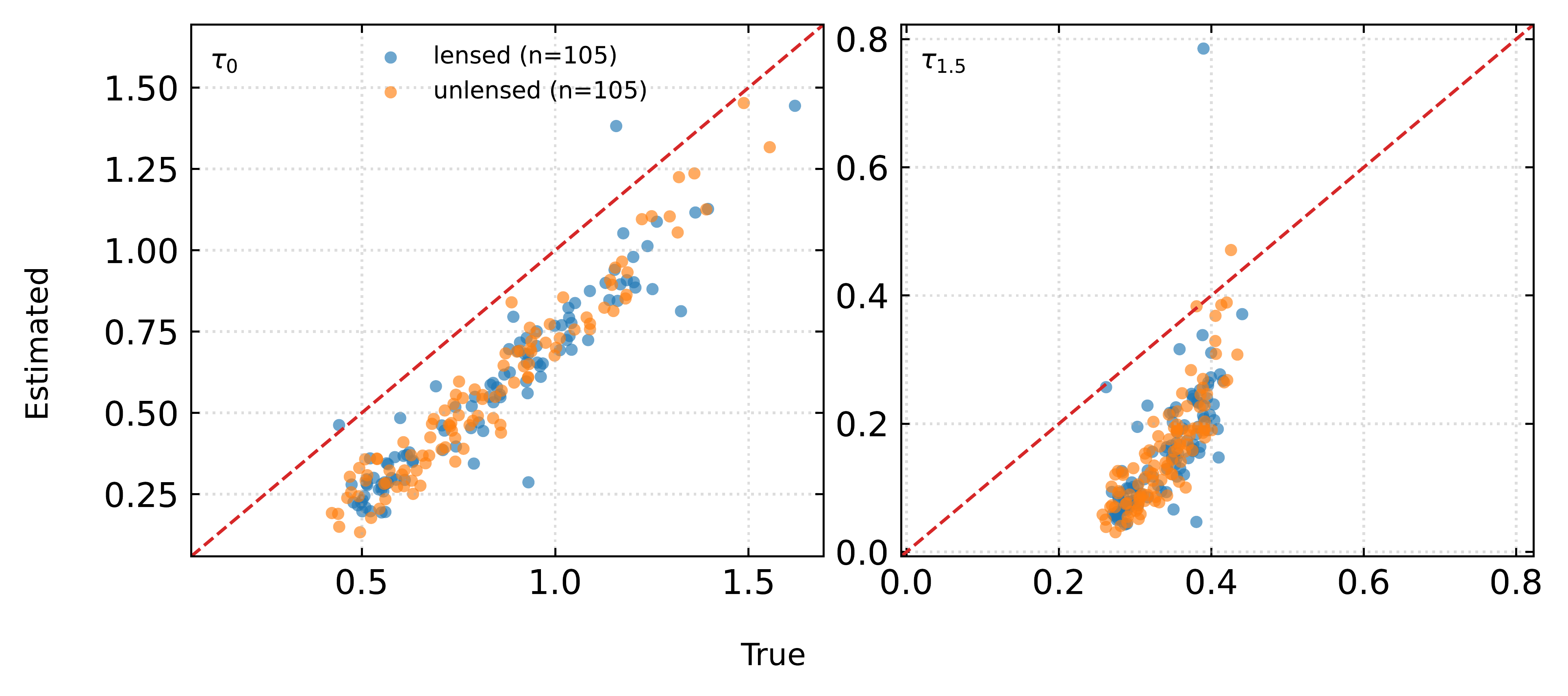}
    \caption{Scatter plot of the PSO parameter estimation results for $\tau_0$ and $\tau_{1.5}$ on the IMRPhenomD data.}
    \label{fig:IMRD_tau}
\end{figure}

During dataset generation, since we constrained the symmetric mass ratio to the range $(0,0.25)$, some of the $\mathcal{M}$ and $q$ values inverted from $\tau_0$ and $\tau_{1.5}$ exceeded this predetermined range and became invalid data. Therefore, we present in Fig.~\ref{fig:IMRD_tau} the PSO parameter estimation results for $\tau_0$ and $\tau_{1.5}$ on the IMRPhenomD data. From the results in the figure, it can be concluded that using the restricted 2-PN template to match IMRPhenomD leads to relatively large parameter estimation errors, with the estimated values of both $\tau_0$ and $\tau_{1.5}$ lying below the ideal results.

\clearpage

\section{}
\begin{figure}[htbp]
    \centering
    \includegraphics[width=\textwidth]{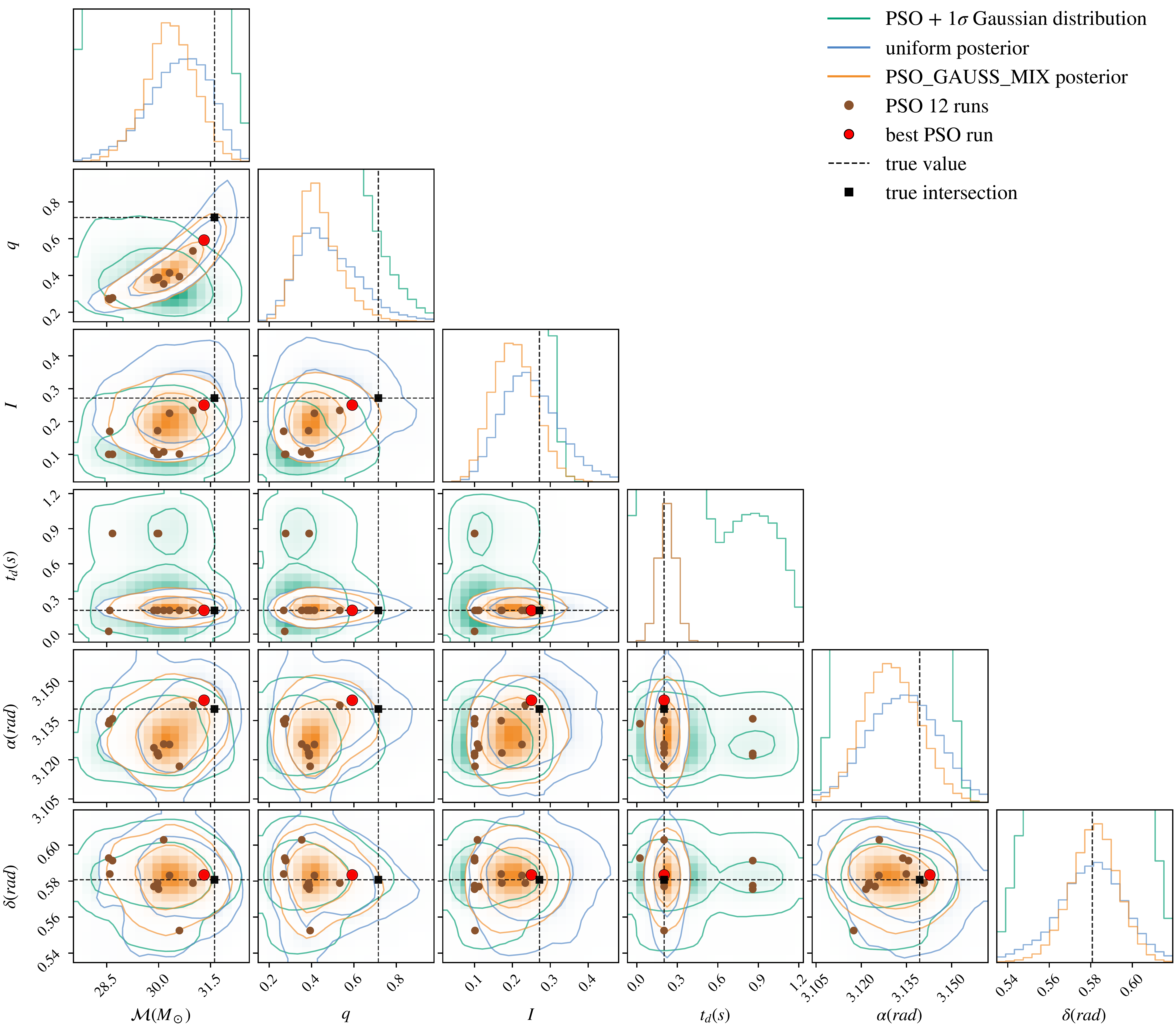}
    \caption{Supplementary comparison including the PSO+$1\sigma$ Gaussian parameter distribution reconstructed from the 12 PSO point estimates. The green contours represent this reconstructed distribution, with the Gaussian width set by the $1\sigma$ scatter of the PSO results, while the blue and orange contours show the nested-sampling posteriors under the uniform and PSO\_GAUSS\_MIX priors, respectively. Other markers follow the same convention as in Fig.~\ref{fig:pso_gauss_mix_corner}.}
    \label{fig:appendix_pso_prior_posterior_comparison}
\end{figure}
\end{document}